\RequirePackage{fix-cm}
\documentclass[twocolumn,epjc3]{svjour3}  
\smartqed  

\RequirePackage{graphicx}
\RequirePackage{dcolumn}
\RequirePackage{bm}
\RequirePackage{color} 
\RequirePackage{xcolor} 
\RequirePackage{color} 
\RequirePackage{graphics}
\RequirePackage{float}
\RequirePackage{bm}        
\RequirePackage{amssymb}   
\RequirePackage{slashed}   
\RequirePackage{amsmath}   
\RequirePackage{verbatim}  
\RequirePackage{subfig}
\RequirePackage{mathptmx}      
\RequirePackage{flushend}
\RequirePackage[numbers,sort&compress]{natbib}
\RequirePackage[colorlinks,citecolor=blue,urlcolor=blue,linkcolor=blue]{hyperref}

\newcounter{YJC}

\journalname{Eur. Phys. J. C}

\begin{document}
\title{Detect anomalous quartic gauge couplings at muon colliders with quantum kernel k-means}

\author{Shuai Zhang\thanksref{e1,addr1,addr2}
        \and 
        Ke-Xin Chen\thanksref{e2,addr1,addr2} 
        \and 
        Ji-Chong Yang\thanksref{cr,e3,addr1,addr2}}

\institute{Department of Physics, Liaoning Normal University, Dalian 116029, China \label{addr1}
           \and
           Center for Theoretical and Experimental High Energy Physics, Liaoning Normal University, Dalian 116029, China \label{addr2}}
\thankstext[$\star$]{cr}{Corresponding author}
\thankstext{e1}{e-mail: 2802368240@qq.com}
\thankstext{e2}{e-mail: 924038358@qq.com}
\thankstext{e3}{e-mail: yangjichong@lnnu.edu.cn}

\date{Received: date / Revised version: date}

\maketitle

\begin{abstract}
In recent years, with the increasing luminosities of colliders, handling the growing amount of data has become a major challenge for future New Physics~(NP) phenomenological research.
In order to improve efficiency, machine learning algorithms have been introduced into the field of high-energy physics.
As a machine learning algorithm, kernel k-means has been demonstrated to be useful for searching NP signals.
It is well known that the kernel k-means algorithm can be carried out with the help of quantum computing, which suggests that quantum kernel k-means~(QKKM) is also a potential tool for NP phenomenological studies in the future.
This paper investigates how to search for NP signals using QKKM.
Taking the $\mu^+\mu^-\to \nu\bar{\nu}\gamma\gamma$ process at a muon collider as an example, the dimension-8 operators contributing to anomalous quartic gauge couplings~(aQGCs) are studied.
The expected coefficient constraints obtained using the QKKM of three different forms of quantum kernels, as well as the constraints obtained by the classical k-means algorithm are presented, and it can be shown that QKKM can help to find the signal of aQGCs.
Comparing the classical k-means anomaly detection algorithm with QKKM, it is indicated that the QKKM is able to archive a better cut efficiency. 
\end{abstract}

\maketitle

\section{\label{sec1}Introduction}

Significant developments have been made in the field of quantum computing.
The development of quantum computers has progressed from theoretical models to practical applications, with quantum processors now capable of performing complex calculations at significantly faster speeds than their classical counterparts. 
Researchers are continuing to advance the frontiers of quantum computing, resulting in groundbreaking developments in hardware, algorithms, and applications~\cite{Arute:2019zxq}.
The capacity to process vast quantities of data at hitherto unattainable speeds renders quantum computing a potentially transformative force across a multitude of disciplines.

Meanwhile, as the large hadron collider~(LHC) experiment enters the post-Higgs discovery era, physicists have begun to work on the search for new physics~(NP) beyond the Standard Model~(SM)~\cite{Ellis:2012zz}.
The search for NP has now become one of the frontiers of high-energy physics~(HEP), who frequently entails the examination of extensive datasets, generated by means of particle collisions or other experimental procedures.
The potential for quantum computing to significantly accelerate data processing and analysis makes it an invaluable tool for advancing the detection of NP signals.
Despite quantum computing is still in the era of noisy intermediate-scale quantum (NISQ) devices~\cite{Preskill:2018jim,Arute:2019zxq,CG2023}, its applications in various aspects of HEP has already been discussed~\cite{Zhu:2024own,Carena:2022kpg,Bauer:2022hpo,Roggero:2018hrn,Roggero:2019myu,Gustafson:2022xdt,Lamm:2024jnl,Carena:2024dzu,Atas:2021ext,Li:2023vwx,Cui:2019sfz,Zou:2021pvl,Georgescu:2013oza,Lamm:2019uyc,Li:2021kcs,Echevarria:2020wct,Jordan:2011ci,Mueller:2019qqj,Chou:2023hcc,Bauer:2019qxa}.

In the phenomenological studies of NP, the SM Effective Field Theory~(SMEFT) is frequently used in recent years.
The SMEFT framework extends the SM to incorporate high-dimensional operators that capture potential NP effects~\cite{Weinberg:1979sa,Grzadkowski:2010es,Brivio:2017vri,Buchmuller:1985jz}.
Research on SMEFT has focused on dimension-6 operators, however, from a phenomenological point of view, the dominant effect in many cases occurs in dimension-8 operators~\cite{Ellis:2018cos,Ellis:2019zex,Ellis:2020ljj,Gounaris:2000dn,Gounaris:1999kf,Senol:2018cks,Fu:2021mub,Degrande:2013kka,Jahedi:2022duc,Jahedi:2023myu,Ellis:2017edi,Gounaris:1999kf}. 
In addition, dimension-8 operators are also important for convex geometric perspective operator spaces~\cite{Bi:2019phv,Zhang:2020jyn,Yamashita:2020gtt}.
As a result, the dimension-8 operators are increasingly being focused on.
For one generation of fermions, there are $895$ different baryon number conserving dimension-8 operators.
It is necessary to conduct a detailed kinematic analysis of each of these operators.
As the number of operators to be considered increases, the efficiency of the process tends to decline.

In order to facilitate the efficient analysis of data, anomaly detection~(AD) machine learning~(ML) algorithms have been employed in previous studies within the field of HEP to search for NP signals~\cite{Zhang:2023khv,Zhang:2023ykh,Dong:2023nir,Zhang:2023yfg,Yang:2022fhw,Yang:2021kyy,Jiang:2021ytz,Vaslin:2023lig,Kuusela:2011aa,Collins:2018epr,Atkinson:2022uzb,Kasieczka:2021xcg,Farina:2018fyg,Cerri:2018anq,vanBeekveld:2020txa,CrispimRomao:2020ucc,Ren:2017ymm,Abdughani:2018wrw,Ren:2019xhp}.
This paper investigates the application of a quantum ML~(QML) algorithm to search for NP, i.e., quantum kernel k-means (QKKM). 
The choice of the k-means algorithm among various ML algorithms is motivated by two factors. 
Firstly, it has been demonstrated to be effective in phenomenological studies of NP~\cite{Zhang:2023yfg}. 
Secondly, the kernel k-means algorithm is compatible with quantum computers.
One potential advantage of QKKM is that, it is pointed out that multi-state swap test on quantum computers can compute inner products of multiple vectors simultaneously~\cite{Liu:2022jsp,Fanizza:2020qjq}.
At the same time, quantum kernels have the potential to transform nonlinear data into linearly separable forms through quantum feature mapping~\cite{Liu:2020lhd}.
This paper aims to compare several different quantum kernel methods, all of which are inner products and have the potential to be accelerated by multi-state swap test.

In this paper, we take the study of dimension-8 operators contributing to anomalous quartic gauge couplings~(aQGCs) as an example.
The sensitivity of the vector boson scattering~(VBS) process to aQGCs and the increasing phenomenological research on aQGCs have led to a wide interest in aQGCs~\cite{Green:2016trm,Chang:2013aya,Anders:2018oin,Zhang:2018shp,Bi:2019phv,Guo:2020lim,Guo:2019agy,Yang:2021pcf,Yang:2020rjt}.
Meanwhile, LHC has been closely following the aQGCs~\cite{ATLAS:2014jzl,CMS:2020gfh,ATLAS:2017vqm,CMS:2017rin,CMS:2020ioi,CMS:2016gct,CMS:2017zmo,CMS:2018ccg,ATLAS:2018mxa,CMS:2019uys,CMS:2016rtz,CMS:2017fhs,CMS:2019qfk,CMS:2020ypo,CMS:2020fqz}. 
With the increasing luminosities on future colliders, the muon colliders can achieve higher energies and luminosities while providing a cleaner experimental environment that is less impacted by the QCD background than the hadron colliders~\cite{Buttazzo:2018qqp,Delahaye:2019omf,Costantini:2020stv,Lu:2020dkx,AlAli:2021let,Franceschini:2021aqd,Palmer:1996gs,Holmes:2012aei,Liu:2021jyc,Liu:2021akf}.
In order to study aQGCs, the process $\mu^+\mu^-\rightarrow \nu\bar{\nu}\gamma\gamma$ at muon colliders is used as a testbed.
This process not only lends itself to the study of aQGCs, a NP operator of widely interest, but also provides a place to validate ML algorithms due to the information lost by final-state neutrinos.
The AD event selection strategy with QKKM is employed to search for aQGCs signals, and expected coefficient constraints, i.e. the projected sensitivities are analyzed. 
It is worth noting that, as an AD algorithm, using the QKKM to search for aQGCs signals does not depend on the studied process.

The rest of the paper is organized as follows.
In Section~\ref{sec2}, a brief introduction to aQGCs and the $\mu^+\mu^-\rightarrow \nu\bar{\nu}\gamma\gamma$ process is given.
The event selection strategy of QKKM is discussed in Section~\ref{sec3}.
Section~\ref{sec4} presents numerical results for the expected coefficient constraints.
Section~\ref {sec5} is a summary of the conclusions.

\section{\label{sec2}aQGCs and the process of \texorpdfstring{$\mu^+\mu^-\rightarrow \nu\bar{\nu}\gamma\gamma$}{muon pair to neutrinos and photons} at the muon colliders}

\begin{figure}[htbp]
\begin{center}
\includegraphics[width=0.8\hsize]{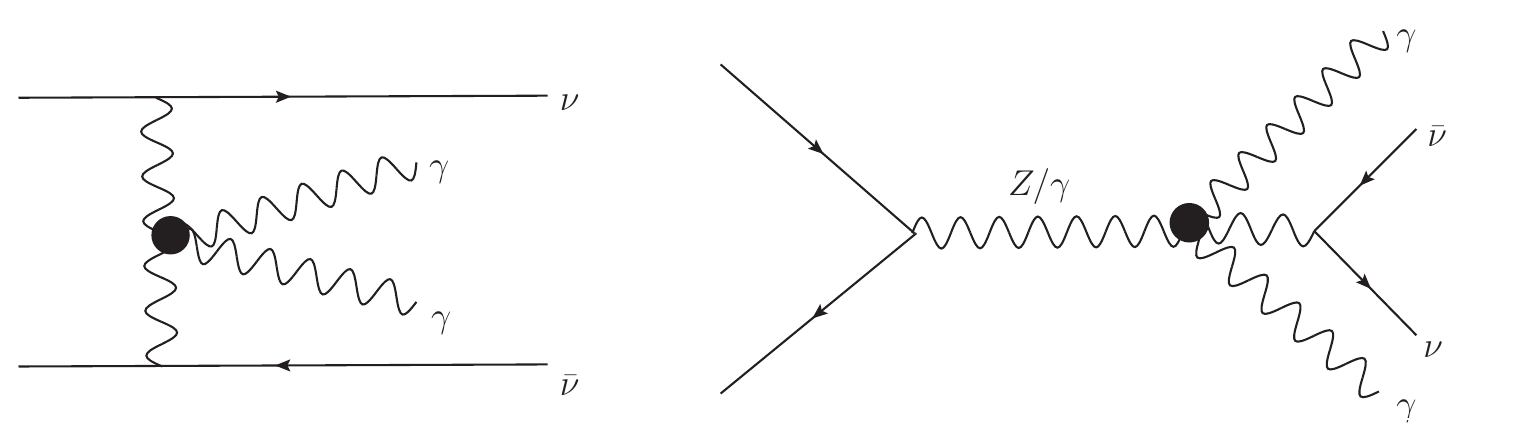}
\caption{\label{fig:Feymans}Typical Feynman diagrams for signal events. At lower energies, the tri-boson process~(in the right panel) dominates, while at higher energies the VBS process~(in the left panel) dominates~\cite{Yang:2020rjt}.}
\end{center}
\end{figure}

\begin{figure}[htbp]
\begin{center}
\includegraphics[width=0.8\hsize]{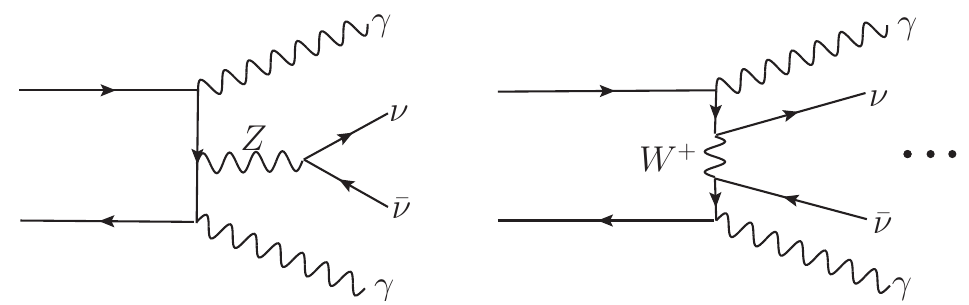}
\caption{\label{fig:Feymanb}Typical Feynman diagrams for background events.}
\end{center}
\end{figure}

The frequently used dimension-8 operators contributing to aQGCs can be classified into three categories, scalar$/$longitudinal operators $O_{S_i}$, mixed transverse and longitudinal operators $O_{M_i}$ and transverse operators $O_{T_i}$, respectively~\cite{Eboli:2006wa,Eboli:2016kko}.
The operators involved in the $\mu^+\mu^-\rightarrow \nu\bar{\nu}\gamma\gamma$ process are $O_{M_i}$ and $O_{T_i}$,
\begin{equation}
\begin{split}
\mathcal{L}_{\rm aQGC} &= \sum_{i}\frac{f_{M_{i}}}{\Lambda^{4}}O_{M,i} + \sum_{j}\frac{f_{T_{j}}}{\Lambda^{4}}O_{T,j},
\end{split}
\label{eq:la}
\end{equation}
where $f_{M_{i}}$ and $f_{T_{j}}$ are dimensionless Wilson coefficients, and $\Lambda$ is the NP energy scale.
The process of $\mu^+\mu^-\rightarrow \nu\bar{\nu}\gamma\gamma$ at the muon collider can be contributed by the operators $O_{M_{0,1,2,3,4,5,7}}$ and $O_{T_{0,1,2,5,6,7}}$, where $O_{M_{0,1,5,7}}$ are not considered due to sensitivities within current coefficient constraints and,
\begin{equation}
\begin{aligned}
O_{M,2} &= \left[B_{\mu\nu}B^{\mu\nu}\right] \times \left[(D_{\beta}\Phi)^{\dagger}D^{\beta}\Phi\right], \\ 
O_{M,3} &= \left[B_{\mu\nu}B^{\nu\beta}\right] \times \left[(D_{\beta}\Phi)^{\dagger}D^{\mu}\Phi\right], \\
O_{M,4} &= \left[(D_{\mu}\Phi)^{\dagger}\widehat{W}_{\alpha\nu}D^{\mu}\Phi\right] \times B^{\beta\nu},\\
O_{T,0} &= \mathrm{Tr}\left[\widehat{W}_{\mu\nu}\widehat{W}^{\mu\nu}\right] \times \mathrm{Tr}\left[\widehat{W}_{\alpha\beta}\widehat{W}^{\alpha\beta}\right],\\
O_{T,1} &= \mathrm{Tr}\left[\widehat{W}_{\alpha\nu}\widehat{W}^{\mu\beta}\right] \times \mathrm{Tr}\left[\widehat{W}_{\mu\beta}\widehat{W}^{\alpha\nu}\right],\\
O_{T,2} &= \mathrm{Tr}\left[\widehat{W}_{\alpha\mu}\widehat{W}^{\mu\beta}\right] \times \mathrm{Tr}\left[\widehat{W}_{\beta\nu}\widehat{W}^{\nu\alpha}\right],\\
O_{T,5} &= \mathrm{Tr}\left[\widehat{W}_{\mu\nu}\widehat{W}^{\mu\nu}\right] \times B_{\alpha\beta}B^{\alpha\beta},\\
O_{T,6} &= \mathrm{Tr}\left[\widehat{W}_{\alpha\nu}\widehat{W}^{\mu\beta}\right] \times B_{\mu\beta}B^{\alpha\nu},\\
O_{T,7} &= \mathrm{Tr}\left[\widehat{W}_{\alpha\mu}\widehat{W}^{\mu\beta}\right] \times B_{\beta\nu}B^{\nu\alpha},
\end{aligned}
\label{eq:operators}
\end{equation}
where $\widehat{W}\equiv \vec{\sigma}\cdot {\vec W}/2$ with $\sigma$ being the Pauli matrices and ${\vec W}=\{W^1,W^2,W^3\}$, $B_{\mu}$ and $W_{\mu}^i$ are $U(1)_{\rm Y}$ and $SU(2)_{\rm I}$ gauge fields, $B_{\mu\nu}$ and $W_{\mu\nu}$ correspond to the field strength tensors, and $D_{\mu}$ is the covariant derivative.
For the process $\mu^+\mu^-\rightarrow \nu\bar{\nu}\gamma\gamma$, the diagrams induced by $O_{M_{i}}$ and $O_{T_{i}}$ operators are shown in the Fig.~\ref{fig:Feymans}, and the Fig.~\ref{fig:Feymanb} shows the typical diagrams in the SM.
Since aQGC decouple from anomalous triple gauge couplings~(aTGCs) starting from dimension-8, we consider only dimension-8 operators.

As a high-energy collider, the muon collider is also considered a gauge boson collider and is therefore well suited to the study of aQGCs. 
The operators that contribute independently to aQGCs and are not related to anomalous triple gauge couplings start at dimension-$8$. 
The muon collider is therefore well suited to study the signal of the dimension-8 aQGCs in the VBS processes. 
Among the many VBS processes, those in which the forward-moving particles are neutrinos have an advantage because there is no need to detect charged particles near the direction of the beams. 
These are processes that contain $WW\to VV$ sub-processes. 
Among these processes, the one in which the final state $VV$ are photons has the least electroweak vertices and is therefore more advantageous, which is the $\mu^+\mu^-\to \nu\bar{\nu}\gamma\gamma$ process that is the focus of this work.
Since there are no indications of higher-dimensional operators yet, we restrict ourselves to focusing only on the projected sensitivities of the aQGCs, ignoring the effects of other SMEFT operators.

As an Effective Field Theory~(EFT), the SMEFT is only valid under the NP energy scale. 
The high center-of-mass~(c.m.) energy achievable at muon colliders offers an excellent opportunity to detect potential NP signals, while at the same time, raises concerns on the validity of the SMEFT. 
Previous studies have extensively employed partial wave unitarity as a criterion for assessing the validity of the SMEFT~\cite{Dong:2023nir,Yang:2021pcf,Guo:2020lim,Yue:2021snv,Fu:2021mub,Yang:2022ilt,Layssac:1993vfp,Corbett:2017qgl,Almeida:2020ylr,Kilian:2018bhs,Kilian:2021whd,Perez:2018kav}.
For the process $W^{-}_{\lambda_{1}}W^{+}_{\lambda_{2}}\rightarrow {\gamma}_{\lambda_{3}}{\gamma}_{\lambda_{4}}$ with $\lambda_{1,2} = \pm 1,0$ and $\lambda_{3,4} = \pm 1$ correspond to the helicities of the vector bosons, in the c.m. frame with z-axis along the flight direction of ${W}^-$ in the initial state, the amplitudes can be expanded as~\cite{Jacob:1959at},
\begin{equation}
\begin{aligned}
\mathcal{M}(W^{-}_{\lambda_{1}}W^{+}_{\lambda_{2}} \rightarrow \gamma_{\lambda_{3}} \gamma_{\lambda_{4}}) &=\\ 8 \pi \sum_{J}(2J+1) \sqrt{1 + \delta_{\lambda_{3}\lambda_{4}}} e^{i(\lambda - \lambda^{'})\phi} d^{J}_{\lambda \lambda^{'}}(\theta) T^{J},
\end{aligned}
\label{eq:M}
\end{equation}
where $\theta$ and $\phi$ are zenith and azimuth angles of $\gamma _{\lambda _3}$, $\lambda = \lambda_{1} - \lambda_{2}$, $\lambda^{'} = \lambda_{3} - \lambda_{4}$
and $d^{J}_{\lambda\lambda^{'}}(\theta)$ is the Wigner D-functions.
The partial wave unitarity bound is $|T^{J}| \leq 2$~\cite{Corbett:2014ora}.

In Refs.~\cite{Guo:2019agy,Yang:2021ukg}, the results of partial wave unitarity bounds on coefficients of the $\gamma\gamma WW$ vertices have been obtained in the study of $\gamma\gamma\to W^+W^-$ VBS process at the LHC. 
The dimension-8 operators contribute to five different $WW\gamma\gamma$ vertices, with each vertex being contributed to by only one operator. Consequently, the partial wave unitarity bounds on coefficients of the $\gamma\gamma WW$ vertices can be directly translated to the partial wave unitarity bounds on operator coefficients for the $WW\to \gamma\gamma$ by assuming one operator at a time.
The strongest partial wave unitarity bounds w.r.t. the $WW\to \gamma\gamma$ process are,
\begin{equation}
\begin{split}
\begin{aligned}
&\left|\frac{f_{M_{2}}}{\Lambda^4}\right| \leq \frac{64\sqrt{2}\pi M_{W}^2{s_{W}^2}}{{s}^2e^2v^2{c_{W}^2}}, &&\left|\frac{f_{M_{3}}}{\Lambda^4}\right| \leq \frac{256\sqrt{2}\pi M_{W}^2{s_{W}^2}}{{s}^2e^2v^2{c_{W}^2}},\\
&\left|\frac{f_{M_{4}}}{\Lambda^4}\right| \leq \frac{128\sqrt{2}\pi M_{W}^2{s_{W}}}{{s}^2e^2v^2{c_{W}}},&&\left|\frac{f_{T_{0}}}{\Lambda^4}\right| \leq \frac{8\sqrt{2}\pi}{{s}^2{s_{W}^2}},\\
&\left|\frac{f_{T_{1}}}{\Lambda^4}\right| \leq \frac{24\sqrt{2}\pi}{{s}^2{s_{W}^2}},&&\left|\frac{f_{T_{2}}}{\Lambda^4}\right| \leq \frac{32\sqrt{2}\pi }{{s}^2{s_{W}^2}},\\
&\left|\frac{f_{T_{5}}}{\Lambda^4}\right| \leq \frac{8\sqrt{2}\pi}{{s}^2{c_{W}^2}},
&&\left|\frac{f_{T_{6}}}{\Lambda^4}\right| \leq \frac{24\sqrt{2}\pi }{{s}^2{c_{W}^2}},\\
&\left|\frac{f_{T_{7}}}{\Lambda^4}\right| \leq \frac{32\sqrt{2}\pi}{{s}^2{c_{W}^2}}.\\
\end{aligned}
\end{split}
\label{eq:fbound}
\end{equation}

\begin{table}[htbp]
\centering
\begin{tabular}{c|c|c} 
\hline
  $\sqrt{s}$&10\;{\rm TeV} &14\;{\rm TeV}\\
  
  \hline

  $\left|f_{M_{2}}/\Lambda^4\right|(\rm TeV^{-4})$ &$9.8\times 10^{-3}$& $7.0\times 10^{-3}$\\
\hline
  $\left|f_{M_{3}}/\Lambda^4\right|(\rm TeV^{-4})$ &$3.9\times 10^{-2}$& $2.8\times 10^{-2}$\\
\hline
  $\left|f_{M_{4}}/\Lambda^4\right|(\rm TeV^{-4})$ &$3.6\times 10^{-2}$& $2.6\times 10^{-3}$\\
\hline
  $\left|f_{T_{0}}/\Lambda^4\right|(\rm TeV^{-4})$  &$1.5\times 10^{-2}$& $1.1\times 10^{-3}$\\
\hline
  $\left|f_{T_{1}}/\Lambda^4\right|(\rm TeV^{-4})$ &$4.6\times 10^{-2}$& $3.3\times 10^{-2}$\\
\hline
  $\left|f_{T_{2}}/\Lambda^4\right|(\rm TeV^{-4})$ &$6.1\times 10^{-2}$& $4.4\times 10^{-2}$\\
\hline
  $\left|f_{T_{5}}/\Lambda^4\right|(\rm TeV^{-4})$ &$4.6\times 10^{-3}$& $3.3\times 10^{-3}$\\
\hline
  $\left|f_{T_{6}}/\Lambda^4\right|(\rm TeV^{-4})$ &$1.4\times 10^{-2}$& $1.0\times 10^{-3}$\\
\hline
  $\left|f_{T_{7}}/\Lambda^4\right|(\rm TeV^{-4})$ &$1.8\times 10^{-2}$& $1.3\times 10^{-3}$\\
\hline
\end{tabular}
\caption{The values of the tightest partial wave unitarity bounds at $\sqrt{s}=10\; \rm TeV$ and $14\;\rm TeV$.}
\label{table:unitaritybounds}
\end{table}
The maximum possible c.m. energy for the subprocess $WW \rightarrow \gamma\gamma$ is identical to the c.m. energy of the process $\mu^+\mu^- \rightarrow \nu\bar{\nu}\gamma\gamma$.  
At muon colliders, we consider two cases of the expected energies, $\sqrt{s} = 10 \; \rm{TeV}$ and $\sqrt{s} = 14 \; \rm{TeV}$~\cite{Palmer:1996gs}, the tightest unitarity bounds are listed in Table~\ref{table:unitaritybounds}.

\begin{table}[htbp]
\centering
\begin{tabular}{c|c|c} 
\hline
 $\sigma(\rm pb)$&$\sqrt{s}(\rm TeV)$&$\sqrt{s}(\rm TeV)$ \\
 &$10$&$14$\\
  \hline
    $\sigma_{\rm SM}$&$1.09\times 10^{-1}$&$1.12\times 10^{-1}$ \\
\hline
$\sigma_{O_{M_2}}$&$2.08\times 10^{-5}$& $1.03\times 10^{-5}$\\
${\sigma}^{int}_{O_{M_2}}$&$-1.89\times 10^{-6}$& $-6.83\times 10^{-7}$\\
\hline
$\sigma_{O_{M_3}}$&$2.35\times 10^{-5}$& $1.17\times 10^{-5}$\\
${\sigma}^{int}_{O_{M_3}}$&$3.42\times 10^{-6}$& $1.29\times 10^{-6}$\\
\hline
$\sigma_{O_{M_4}}$ &$2.14\times 10^{-5}$& $1.08\times 10^{-5}$\\
${\sigma}^{int}_{O_{M_4}}$ &$-1.78\times 10^{-6}$& $-6.44\times 10^{-7}$ \\
\hline
$\sigma_{O_{T_0}}$&$2.10\times 10^{-4}$& $2.05\times 10^{-4}$\\
${\sigma}^{int}_{O_{T_0}}$ &$1.02\times 10^{-4}$& $8.51\times 10^{-5}$ \\
\hline
$\sigma_{O_{T_1}}$&$2.11\times 10^{-4}$& $2.13\times 10^{-4}$\\
${\sigma}^{int}_{O_{T_1}}$ &$6.92\times 10^{-5}$& $5.79\times 10^{-5}$   \\
\hline
$\sigma_{O_{T_2}}$ &$2.17\times 10^{-4}$& $2.07\times 10^{-4}$\\
${\sigma}^{int}_{O_{T_2}}$ &$1.22\times 10^{-4}$& $1.01\times 10^{-4}$ \\
\hline
$\sigma_{O_{T_5}}$ &$2.08\times 10^{-4}$& $2.13\times 10^{-4}$\\
${\sigma}^{int}_{O_{T_5}}$ &$2.09\times 10^{-5}$& $1.71\times 10^{-5}$ \\
\hline
$\sigma_{O_{T_6}}$ &$2.07\times 10^{-4}$& $2.03\times 10^{-4}$\\
${\sigma}^{int}_{O_{T_6}}$ &$7.16\times 10^{-5}$& $6.25\times 10^{-5}$  \\
\hline
$\sigma_{O_{T_7}}$ &$2.11\times 10^{-4}$& $2.21\times 10^{-4}$\\ ${\sigma}^{int}_{O_{T_7}}$ &$1.37\times 10^{-4}$& $1.71\times 10^{-4}$ \\
\hline
\end{tabular}
\caption{At $\sqrt{s}=10\;\rm TeV$ and $14\;\rm TeV$, the contribution from the SM, the operators $O_{M_{2,3,4}}$ and $O_{T_{0,1,2,5,6,7}}$, and the interference between the SM and NP.}
\label{table:interference}
\end{table}

K-means AD algorithm can be utilized to address interference~\cite{Zhang:2023yfg}.
However, in this paper, due to the limited computational resources, we do not consider the interference terms for simplicity. 
The contribution of the SM~(denoted as $\sigma _{\rm SM}$), the NP~(denoted as $\sigma _{O_X}$ where $X$ is the name of operator), and the interference between the SM and NP~(denoted as $\sigma ^{int}_{O_X}$) for different operators when the coefficients are the upper bounds in Table~\ref{table:unitaritybounds} at $\sqrt{s} = 10 \;\rm{TeV}$ and $14 \;\rm{TeV}$ are listed in Table~\ref{table:interference}.
We only study operators where the cross-sectional ratio of the interference term to the NP contribution is less than $O_{M_3}$~(for $O_{M_3}$, $\sigma ^{int}_{O_{M_i}}/\sigma _{O_{M_i}}$ is $14.56\%$ at $\sqrt{s}=10\;{\rm TeV}$).
That is, we focus on the operators $O_{M_{2,3,4}}$ and $O_{T_{5}}$.

The smaller the coefficient, the more important the interference terms are.
From Table~\ref{table:unitaritybounds}, it can be seen that the unitarity bounds are not yet at a level where the interference terms become dominant.
Dimensional analysis indicates that $\sigma ^{int}_{NP}\sim sf_X/\Lambda ^4$ and $\sigma _{NP}\sim s^3\left(f_X/\Lambda ^4\right)^2$, when $\sigma ^{int}_{NP}\approx \sigma _{NP}$, $f_X/\Lambda ^4\leq s^{-2}$.
At $\sqrt{s}=10\;{\rm TeV}$, the rough estimation is that, the interference terms become important when $f_X/\Lambda ^4\leq 10^{-4}\;{\rm TeV}^{-4}$.
From the numerical results that follow, the interference term can be ignored in the following sections. 
This is mainly due to the fact that the constraints on $f_X$ are not small enough. 
Moreover, we mainly consider the $O_{M,T}$ operators, where the dominant contributions come from the scattering of the transversely polarized $W$, which is different from the the case of the SM where the contribution of scattering of longitudinally polarized $W$ dominants, and thus the interference terms are suppressed.

The relative contributions of the VBS processes to the annihilation process are contingent upon the specific process under consideration~\cite{Chen:2022yiu,Han:2024gan}, particularly in scenarios where the interference term is important. 
For the process $\mu^+\mu^-\to \nu\bar{\nu}\gamma\gamma$, a comparison of the contributions of the VBS and tri-boson induced by $O_{T_5}$ when the interference term is not considered is provided in Ref.~\cite{Yang:2020rjt}. 
In this case, the VBS contribution exceeds that of the tri-boson at approximately $\sqrt{s}=5\;{\rm TeV}$.
For the $O_M$ operators, the contributions are presented in the ~\ref{secap}. 
The tri-boson contribution is at the next order of $M_Z^2/s$ compared to the VBS. 
Consequently, it can be expected that at a smaller $s$ compared with $O_{T_5}$, the VBS contribution will dominant. 
This also explains the focus on the $O_M$ operators in this study, since the annihilation processes usually have larger interferences~\cite{Chen:2022yiu,Han:2024gan,Yang:2020rjt} which is ignored.

\section{\label{sec3}The event selection strategy of QKKM}

As the luminosities of future colliders continue to increase, so does the quantity of data that must be processed which presents a significant challenge to conventional computing.
Nevertheless, forecasts by IBM, Google, and IonQ indicate that within the next decade, it will become feasible to execute practical computational tasks using quantum computers with thousands of qubits.
This coincides with the high luminosity upgrade of the LHC~\cite{Apollinari:2015wtw} and future colliders such as muon colliders.
In recent years, numerous QML algorithms have been the subject of study within the field of HEP, such as QSVM, quantum variational classifiers etc~\cite{Guan:2020bdl,Wu:2021xsj,Wu:2020cye,Terashi:2020wfi,Zhang:2023ykh}

The inherent properties of coherence and entanglement in quantum systems endow quantum computers with powerful parallel computing capabilities.
The main motivation for this study lies in its potential future applications.
In contrast to classical computers, quantum computers are capable of storing and processing a greater quantity of data simultaneously. 
It is conceivable that in the future, the data that must be processed may originate directly from quantum computers. 
In addition, quantum computer can implement kernel functions that are difficult to achieve with classical computers~\cite{Havlicek:2018nqz,Liu:2020lhd,Sherstov:2020qax}.
In this section, we use the QKKM to verify its feasibility in searching for NP.

\subsection{\label{sec3.1}Data preparation}

In order to investigate the QKKM, the events are generated using coefficients that correspond to the upper bounds of the partial wave unitarity constraints.
The Monte Carlo~(MC) simulation is performed using the \verb"MadGraph5@NLO" toolkit~\cite{Alloul:2013bka,Alwall:2014hca,Christensen:2008py,Degrande:2011ua}, while the muon collider-like detector simulation is conducted with the \verb"Delphes"~\cite{deFavereau:2013fsa} software. 
The analysis of the signals and the background is performed with the \verb"MLAnalysis"~\cite{MLAnalysis}.
To avoid infrared divergences, we use the basic cut as the default setting.
The cut relevant to infrared divergences are,
\begin{equation}
\begin{split}
&p_{T,\gamma} > 10\;{\rm GeV},\;\; |\eta _{\gamma}| < 2.5, \;\; \Delta R_{\gamma\gamma} > 0.4,
\end{split}
\label{eq.standardcuts}
\end{equation}
where $p_{T,\gamma}$ and $\eta _{\gamma}$ are the transverse momentum and pseudo-rapidity for each photon, respectively, $\Delta R_{\gamma\gamma}=\sqrt{\Delta \phi_{\gamma\gamma} ^2 + \Delta \eta_{\gamma\gamma}^2}$ where $\Delta \phi_{\gamma\gamma}$ and $\Delta \eta_{\gamma\gamma}$ are differences between the azimuth angles and pseudo-rapidities of two photons.
The signal events for are generated with one operator at a time.

\begin{table}[htbp]
\centering
\begin{tabular}{c|c|c|c|c|c|c} 
\hline
  & $ p^{1}$& $ p^{2}$ & $p^{3}$ & $p^{4}$& $p^{5}$& $ p^{6}$ \\
 \hline
 observables & $E_{\gamma_{1}}$& $p^x_{\gamma_{1}}$&$p^y_{\gamma_{1}}$&$p^z_{\gamma_{1}}$ & $E_{\gamma_{2}}$&$p^x_{\gamma_{2}}$\\
  \hline
   & $ p^{7} $ & $p^{8} $ & $ p^{9}$& $p^{10}$& $ p^{11}$& $ p^{12}$ \\  
  \hline
  observables & $p^y_{\gamma_{2}}$& $p^z_{\gamma_{2}}$&$E_{miss}$ & $p^x_{miss}$ & ${p}^y_{miss}$&${p}^z_{miss}$\\
  \hline
\end{tabular}
\caption{The events are mapped into points in a 12-dimensional space, and a point is denoted as $\vec{p}$. The components of $\vec{p}$ and the corresponding observables are listed.}
\label{table:observable}
\end{table}

In order to collect the features, we require that the final state contains at least two photons~(which is denoted as $N_{\gamma}$ cut). 
At a lepton collider, conservation of momentum can be employed to ascertain the full set of missing momentum components.
In this paper, we choose the components of the four-momenta of the two hardest photons~(the hardest photon is denoted as $\gamma _1$ and the second hardest photon is denoted as $\gamma _2$, respectively) and the missing momentum.
These observables form a 12-dimensional vector denoted by $p$, of which the components $p^{i}$ are listed in Table ~\ref{table:observable}. 

\begin{table}[htbp]
\centering
\begin{tabular}{c|c|c|c} 
\hline
  $\sqrt{s} ({\rm TeV})$ & $\bar p^{1} ({\rm GeV})$& $ \bar p^{2} ({\rm GeV})$ & $ \bar p^{3} ({\rm GeV}) $ \\ 
  \hline
 $10$ & $3.500\times 10^{2}$& $6.858$&$3.256$\\
  \hline
 $14$ &$4.140\times 10^{2}$& $-1.753$& $-0.521$\\
  \hline
  $\sqrt{s} ({\rm TeV})$ & $ \bar p^{4} ({\rm GeV})$& $ \bar p^{5} ({\rm GeV})$& $ \bar p^{6}({\rm GeV}) $  \\ 
  \hline
 $10$  &$-15.374$&$88.006$& $0.231$\\
  \hline
 $14$ &$-2.280$&$96.030$& $1.789$\\
  \hline
  $\sqrt{s} ({\rm TeV})$ & $ \bar p^{7}({\rm GeV})$& $ \bar p^{8}({\rm GeV}) $ & $ \bar p^{9}({\rm GeV})$\\ 
  \hline
 $10$&$-0.440$ & $1.236$ &$9.141\times 10^{3}$\\
  \hline
 $14$ &$0.454$& $2.962$&$1.310\times 10^{4}$\\
  \hline
  $\sqrt{s} ({\rm TeV})$ & $ \bar p^{10}({\rm GeV})$& $\bar p^{11}({\rm GeV})$ & $ \bar p^{12}({\rm GeV})$\\ 
  \hline
 $10$&$-14.133$ & $-5.718$ &$27.888$\\
  \hline
 $14$ &$1.211\times 10^{-2}$&$0.210$& $-1.485$\\
  \hline
\end{tabular}
\caption{The mean values of $j$-th feature $p^j$
over the SM training dataset at $10\;{\rm TeV}$ and $ 14\;{\rm TeV}$, respectively.}
\label{table:mean}
\end{table}

\begin{table}[htbp]
\centering
\begin{tabular}{c|c|c|c} 
\hline
  $\sqrt{s} ({\rm TeV})$ & $ z^{1} ({\rm GeV})$& $  z^{2} ({\rm GeV})$ & $  z^{3} ({\rm GeV}) $ \\ 
  \hline
 $10$ & $5.262\times 10^{2}$& $2.022\times 10^{2}$&$1.906\times 10^{2}$\\
  \hline
 $14$ &$6.884\times 10^{2}$& $2.338\times 10^{2}$& $2.727\times 10^{2}$\\
  \hline
  $\sqrt{s} ({\rm TeV})$ & $  z^{4} ({\rm GeV})$& $  z^{5} ({\rm GeV})$& $  z^{6}({\rm GeV}) $  \\ 
  \hline
 $10$  &$5.673\times 10^{2}$&$1.462\times 10^{2}$& $6.312\times 10^{1}$\\
  \hline
 $14$ &$7.185\times 10^{2}$&$1.839\times 10^{2}$& $8.440\times 10^{1}$\\
  \hline
  $\sqrt{s} ({\rm TeV})$ & $  z^{7}({\rm GeV})$& $  z^{8}({\rm GeV}) $ & $  z^{9}({\rm GeV})$\\ 
  \hline
 $10$&$6.628\times 10^{1}$ & $1.440\times 10^{2}$ &$1.224\times 10^{3}$\\
  \hline
 $14$ &$8.925\times 10^{1}$& $1.672\times 10^{2}$&$1.596\times 10^{3}$\\
  \hline

  $\sqrt{s} ({\rm TeV})$ & $  z^{10}({\rm GeV})$& $ z^{11}({\rm GeV})$ & $  z^{12}({\rm GeV})$\\ 
  \hline
 $10$&$4.078\times 10^{2}$ & $3.782\times 10^{2}$ &$1.145\times 10^{3}$\\
  \hline
 $14$ &$4.618\times 10^{2}$&$5.407\times 10^{2}$& $1.456\times 10^{3}$\\
  \hline

\end{tabular}
\caption{The standard deviations values of $j$-th feature $z^j$ over the SM training dataset at c.m. energies $\sqrt{s} = 10\;{\rm TeV}$ and $ 14\;{\rm TeV}$, respectively. }
\label{table:standard}
\end{table}

Before training, the dataset is standardized by using z-score standardization~\cite{Donoho_2004},
\begin{equation}
\begin{split}
&x_{i}^{j} = \frac{{p_{i}^{j} - \bar{p}^{j}}}{z^{j}},
\end{split}
\label{eq:zscore}
\end{equation}
where $\bar{p}^{j}$ and $z^j$ represent the mean value and standard deviation of the $j$-th feature over the SM training datasets. 
The values of $\bar{p}^{j}$ and $z^j$ at different c.m. energies are listed in Tables~\ref{table:mean} and \ref{table:standard}, respectively.

\subsection{\label{sec3.2}Using QKKM to search for aQGCs}

In this paper, kernel k-means algorithm is used to replace the k-means algorithm. 
The number of clusters in the kernel k-means is denoted as $l$.
The steps to implement QKKM are shown as follows~\cite{Zhang:2023yfg},
\begin{enumerate}
\item Use the quantum circuits to calculate the kernel matrices. 
\item Compute the centroids of the $l$ clusters by substituting the precomputed kernel matrices into the \verb"tslearn" package~\cite{JMLR:v21:20-091}.
\item  Repeat steps 2 for $m$ times.
\item Calculate the anomaly score for each point, i.e., the distance~(denoted by $d$) from the point to the centroid with the same $l$ value~(cluster assignment) as the point.
\item Calculate the average anomaly score~(denoted by $\bar{d}$) over $m$ iterations.
\item Use $\bar{d}>{d}_{th}$ as a cut to select the events.
\end{enumerate}
The only difference between this paper and Ref.~\cite{Zhang:2023yfg} is that the calculation of kernel matrix is implemented using a quantum circuit.

\begin{figure*}[htpb]
\begin{center}
\includegraphics[width=1\hsize]{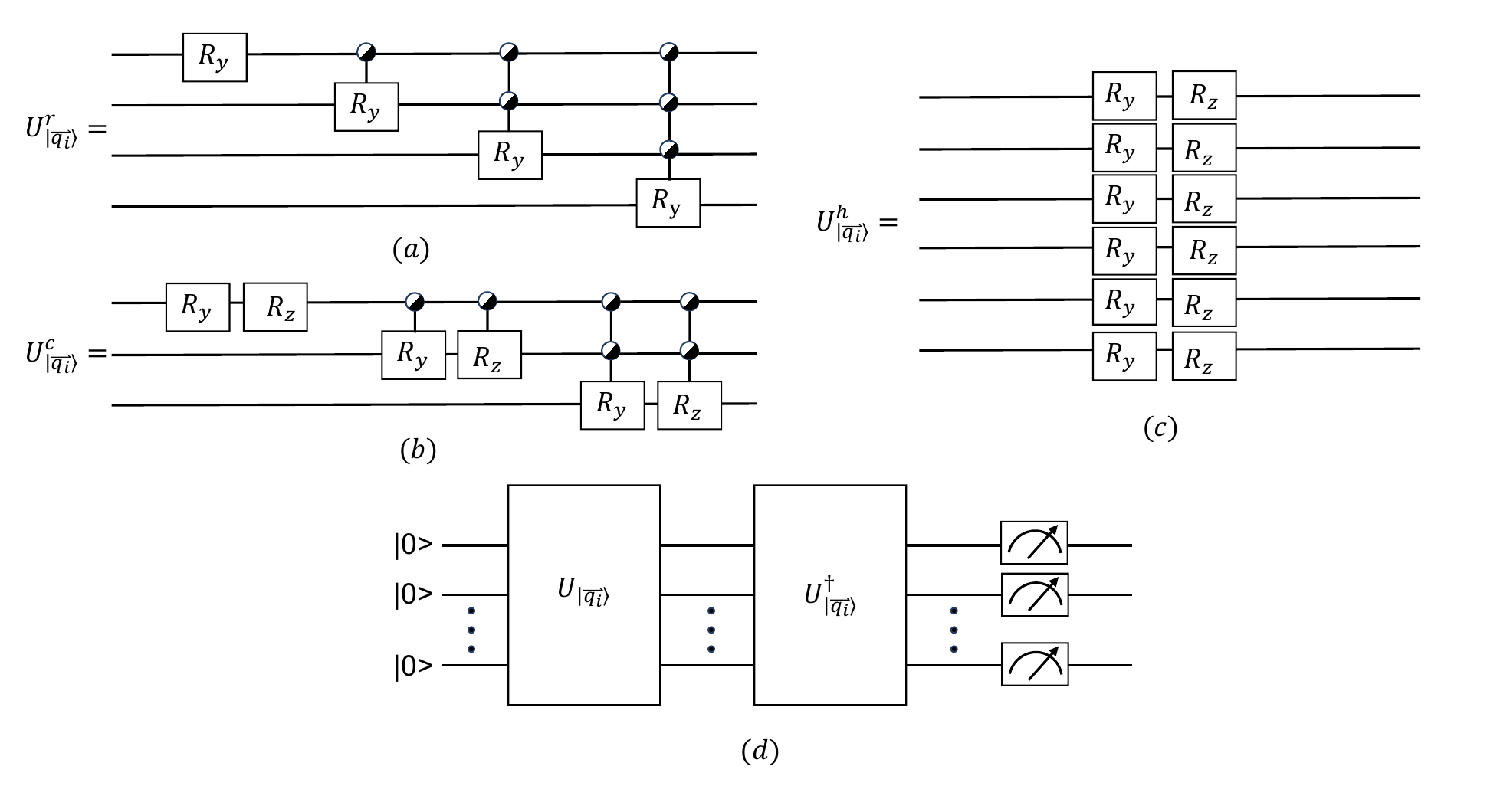}
\caption{\label{fig:circuit}
The circuits used in this paper.
The circuit for amplitude encoding using uniform rotation gates $FR_y$ is depicted in (a). 
The circuit for amplitude encoding using uniform rotation gates $FR_y$ and $FR_z$ is depicted in (b).
The circuit for amplitude encoding using single qubit gates $R_y$ and $R_z$ is depicted in (c).
The circuit to calculate $|\langle \vec{q}_i|\vec{q}_j\rangle|$ is depicted in (d) where the probability of the outcome $|0...00\rangle$ in the measurement is $|\langle \vec{q}_i|\vec{q}_j\rangle|^2$.}
\end{center}
\end{figure*} 

\begin{table}[htbp]
\centering
\begin{tabular}{c|c|c} 
\hline
kernel & single qubit gate & CNOT gate\\
\hline
real vector kernel&$29$&$26$\\
\hline
complex vector kernel&$21$&$16$\\
\hline
hardware-efficient kernel&$6$&$0$\\
\hline
\end{tabular}
\caption{The number of quantum gates to calculate the distance in the cases of three different kernels.
The number of single qubit gates are counted as the number of combined $U$ gates.}
\label{table:gatenumber}
\end{table}

Due to limited computational resources, the 5000 SM events are selected for training.
To map the vector to a quantum state, we use both real and complex vector mapping, i.e., the quantum state presenting the event is denoted as,
\begin{equation}
\begin{split}
|\vec{q}\rangle &= \frac{1}{\sqrt{\sum _i x_i^2 + 1}}(|0\rangle+\sum _{n=1} x_n |n\rangle), \\
|\vec{q}\rangle &= \frac{1}{\sum_{i} x_{i}^2+1}[(1+ x_{1}\rm{i})|0\rangle + \sum_{n=1}{(x_{2n}+x_{2n+ 1}\rm{i}) }|n\rangle], \\
\end{split}
\label{eq.q}
\end{equation}
where $n$ is the digital representing of a state, $x_i$ is the $i$-th component of $\vec{x}$ defined in Eq.~(\ref{eq:zscore}), and $x_{i>12}=0$.
Using Eq.~(\ref{eq.q}), the length of $\vec{x}$ is also encoded.
An $q$ qubit state can encode a vector with $2^{q+1}-2$ degrees of freedom.
Using Eq.~(\ref{eq.q}), a complex vector can be encoded using three qubits, and a real vector can be encoded using four qubits.
In this paper, we use amplitude encode which is denoted as $U^{c}_{\vec{q}}$, such that $U^{c}_{\vec{q}}|0\rangle = |\vec{q}\rangle$.
The amplitude encode of Eq.~(\ref{eq.q}) is implemented with the help of uniform rotation gates $FR_y$ and $FR_z$~\cite{Mottonen:2004vly} as shown in Fig.~\ref{fig:circuit}.~(a) and (b).
Apart from Eq.~(\ref{eq:zscore}), we also try the hardware-efficient encoding~\cite{Wu:2020cye,Fadol:2022umw,Havlicek:2018nqz,Bravo-Prieto:2019kld,Kandala:2017vok,Park:2024rim}.
The hardware-efficient encoding usually consists of multiple layers.
Each layer consists of single qubit $R_{y,z}$ gates with the variables as degrees to be rotated, and the layers are separated by CNOT or controlled-Z gates connecting different qubits.
The hardware-efficient encoding is difficult to be implemented using classical computers.
Since there are only $12$ variables, a single layer is sufficient, necessitating the use of $6$ qubits, as shown in Fig.~\ref{fig:circuit}.~(c).
The swap test can only calculate the absolute value of the inner product, therefore one cannot distinguish between inner product results of $+1$ and $-1$. 
To overcome this limitation, five cases are tested, i.e., we assign the angles to be rotated as $x$, $x/2$, $x/4$, $x/6$ and $x/8$, and $x/8$ yields the best performance.
In the following, only the results with $x/8$ are shown.
The number of gates used in the three types of encodings are listed in Table~\ref{table:gatenumber}.

To calculate the centroids, the distance needs to be defined, which is,
\begin{equation}
\begin{split}
&d(\vec{q}_i,\vec{q}_j)=\sqrt{1- k(\vec{q}_i,\vec{q}_j)},\\
\end{split}
\label{eq.distance}
\end{equation}
where $k$ is the kernel function.
The kernel function is $k(\vec{q}_i, \vec{q}_j) = \left|\langle \vec{q}_i | \vec{q}_j\rangle\right|$, which can be calculated using a circuit shown in Fig.~\ref{fig:circuit}.~(d).
The probability of the outcome $|0\ldots 00\rangle$ in the measurements is $\left|\langle\vec{q}_i|\vec{q}_j\rangle\right|^2$ in the circuit shown in Fig.~\ref{fig:circuit}.~(d).
The calculation of the kernel matrix is implemented using \verb"QuEST"~\cite{Jones:2019knd}. 
The measurement is repeated for $1000$ times for each inner product.

\begin{figure}[htbp]
\begin{center}
\includegraphics[width=0.48\hsize]{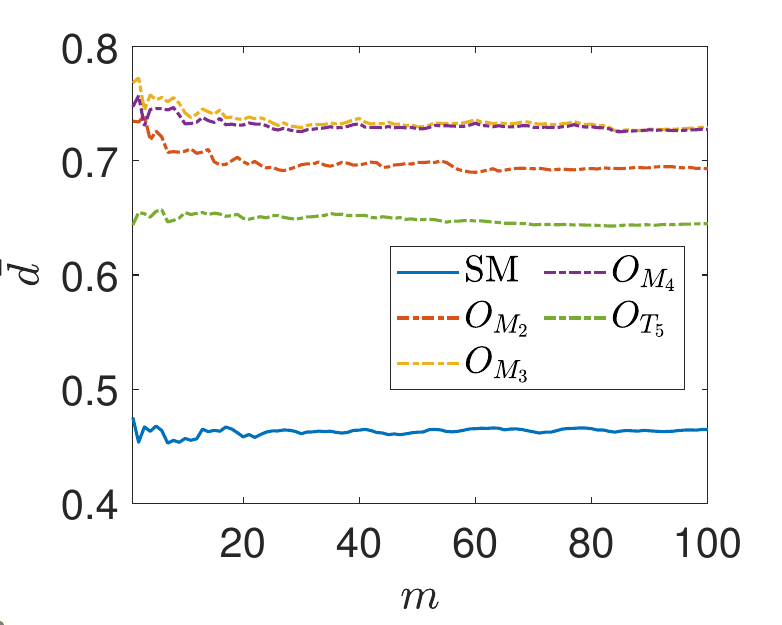}
\includegraphics[width=0.48\hsize]{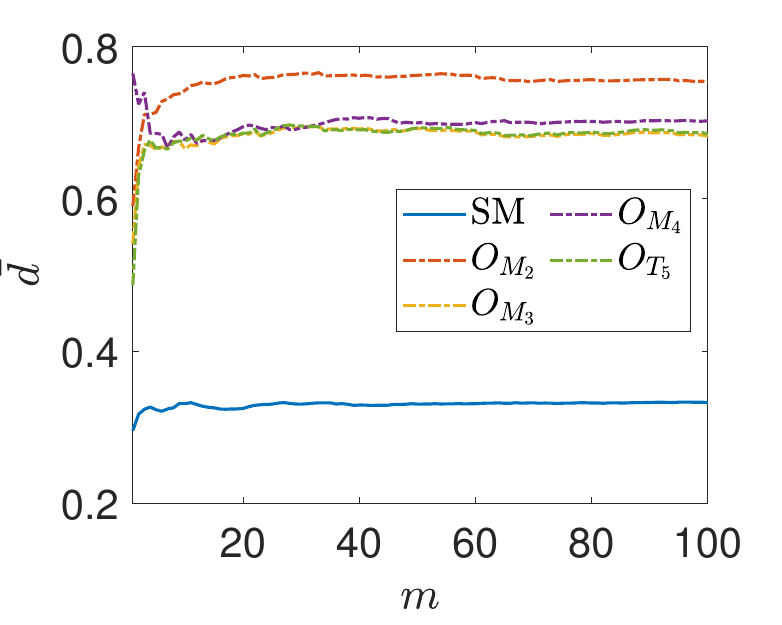}
\includegraphics[width=0.48\hsize]{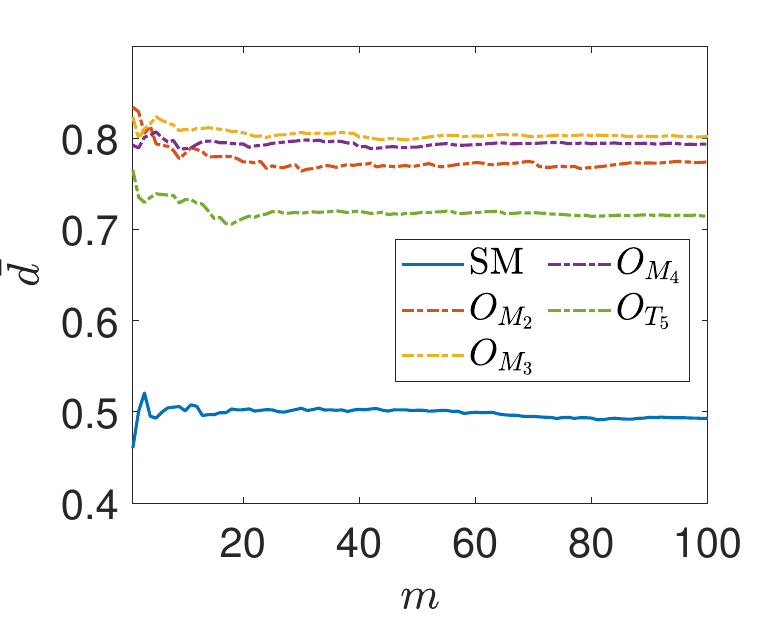}
\includegraphics[width=0.48\hsize]{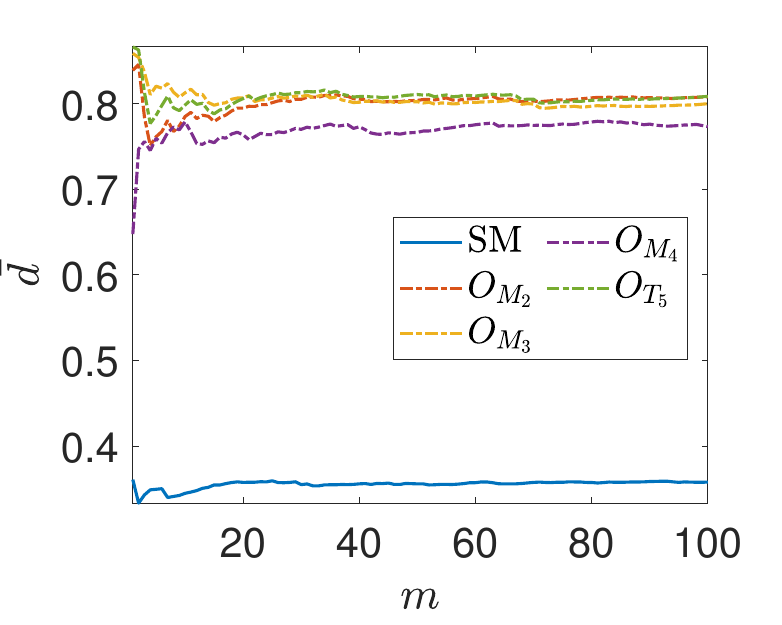}
\includegraphics[width=0.48\hsize]{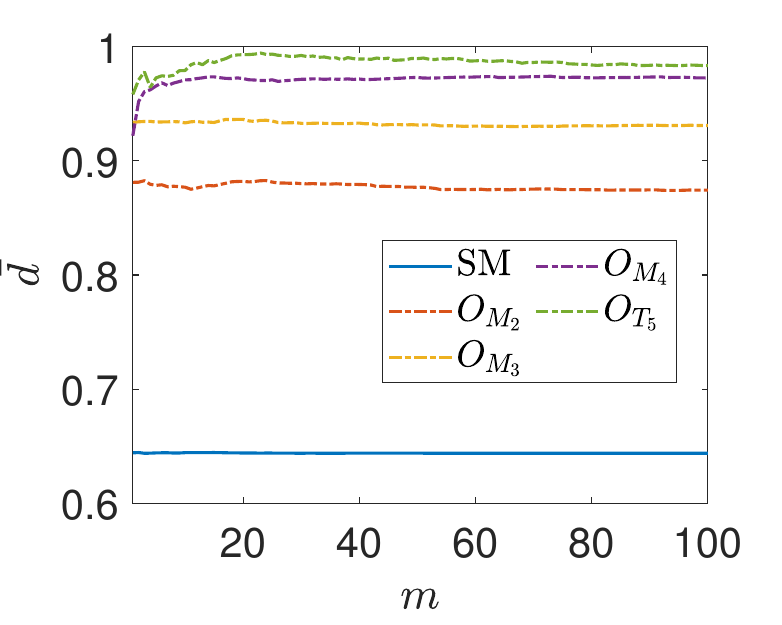}
\includegraphics[width=0.48\hsize]{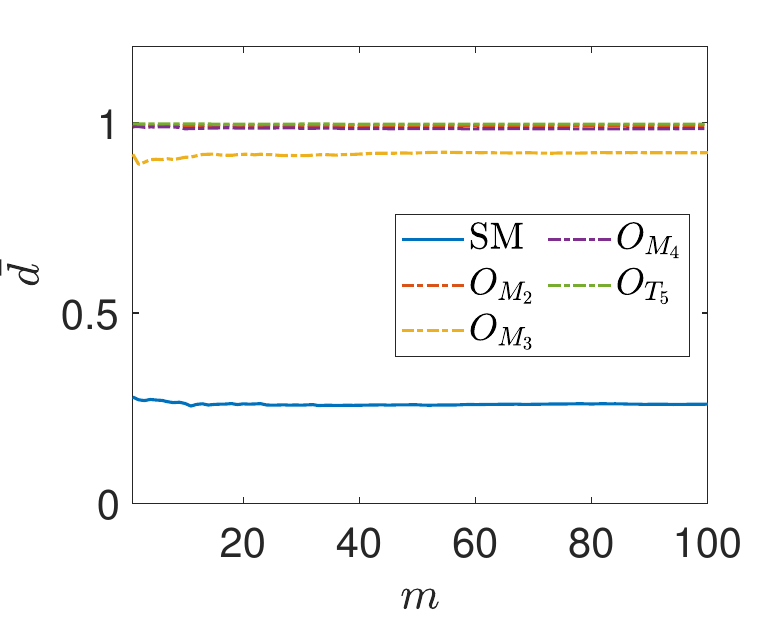}
\caption{\label{fig:md}Taking one collision event from the SM and NP contributions as examples, $\bar{d}$ as function of $m$ at $l=30$. At $10\;\rm TeV$~(the first column) and $14\;\rm TeV$~(the second column), these diagrams correspond to real vector kernel~(the first row), complex vector kernel~(the second row) and hardware-efficient kernel~(the third row), respectively.
We find that $\bar{d}$ converges rapidly with increasing $m$.}
\end{center}
\end{figure}

Due to the random nature of the k-means algorithm, the results of the centroids are not unique. 
To circumvent this issue, the process is repeated $m$ times, where $m$ is a tunable parameter. 
At $\sqrt{s}=10\;{\rm TeV}$ and $14\;{\rm TeV}$, one event is selected from the SM background and $O_{M_{2,3,4}}$ and $O_{T_{5}}$ signals, respectively, and $\bar{d}$ is shown as a function of $m$ at $l = 30$ in Fig.~\ref{fig:md}.
We found that $\bar{d}$ rapidly converges with increasing $m$, and when $m = 100$, the value of $\bar{d}$ begins to stabilize. 
Theoretically, the value of $m$ can be further increased to reduce the relative statistical error of $\bar{d}$.
However, due to limited computational power, we use $m = 100$ in this paper.

\begin{figure}[htbp]
\begin{center}
\includegraphics[width=0.48\hsize]{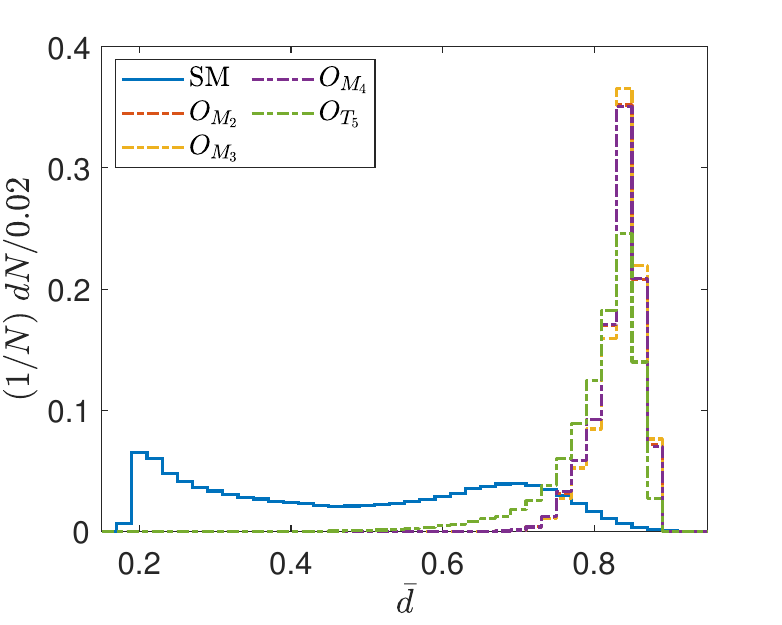}
\includegraphics[width=0.48\hsize]{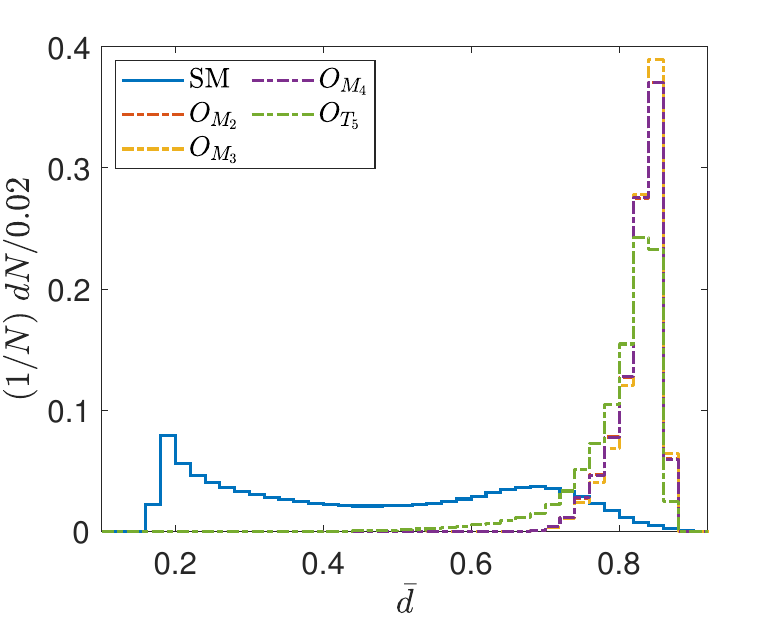}
\includegraphics[width=0.48\hsize]{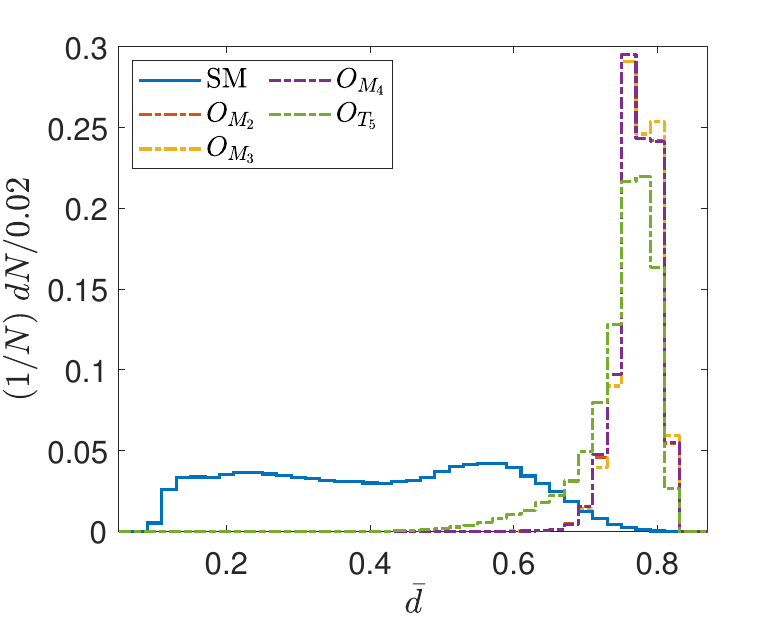}
\includegraphics[width=0.48\hsize]{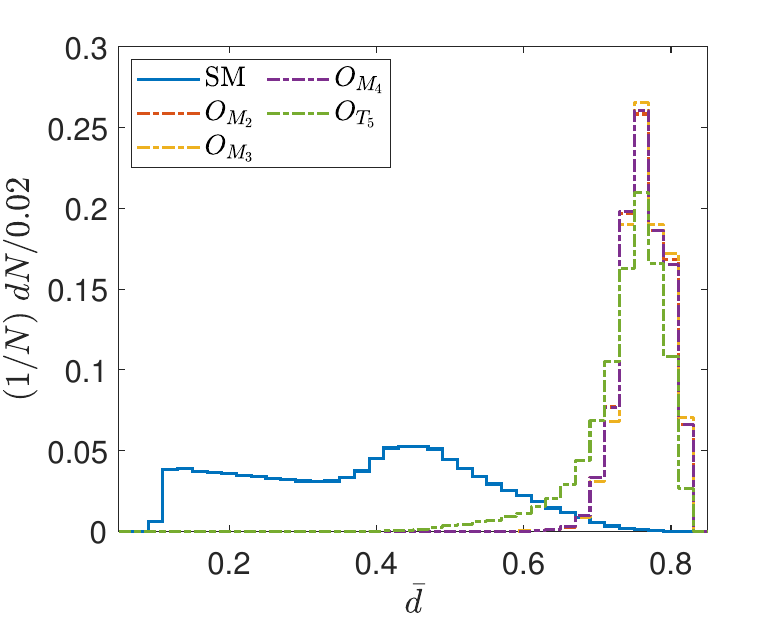}
\includegraphics[width=0.48\hsize]{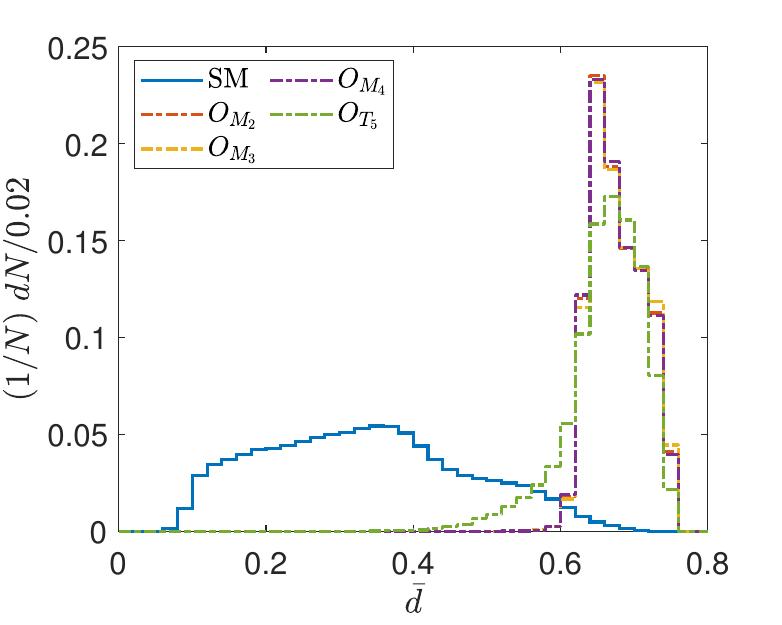}
\includegraphics[width=0.48\hsize]{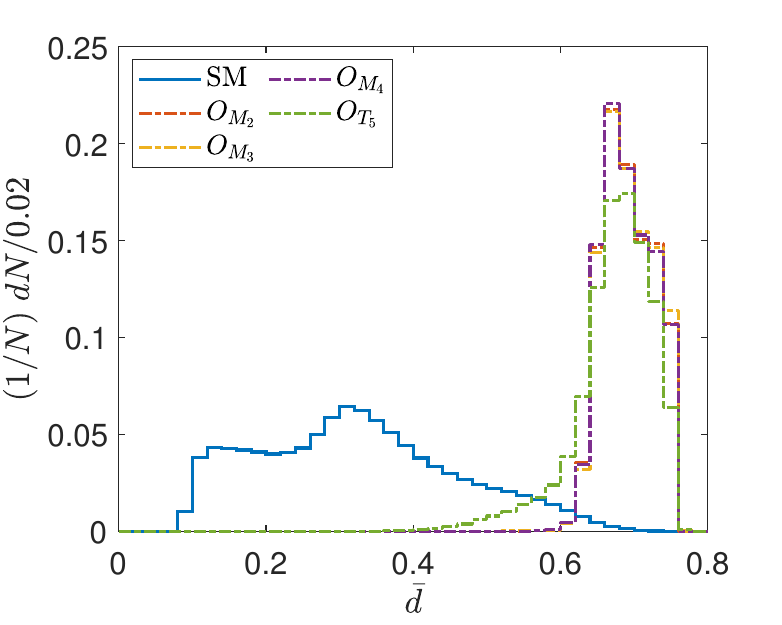}
\caption{\label{fig:l}For the complex vector kernel, the normalized distribution of anomaly score $\bar{d}$ when $l=2$~(the first row), $10$~(the second row), and $30$~(the third row), at $10\;{\rm TeV}$~(the first column) and $14\;{\rm TeV}$~(the second column) for the SM and $O_{M_{2,3,4}}$ and $O_{T_{5}}$ induced contribution.}
\end{center}
\end{figure}

\begin{figure}[htbp]
\begin{center}
\includegraphics[width=0.48\hsize]{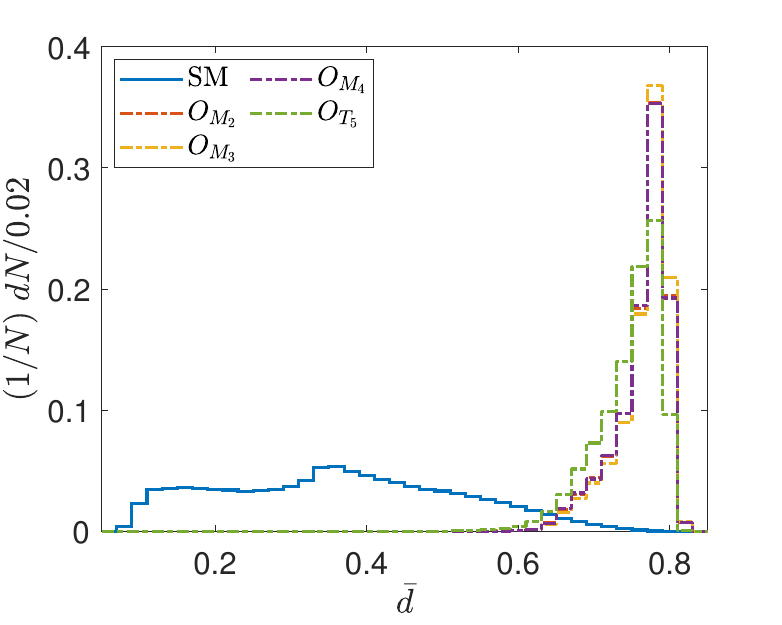}
\includegraphics[width=0.48\hsize]{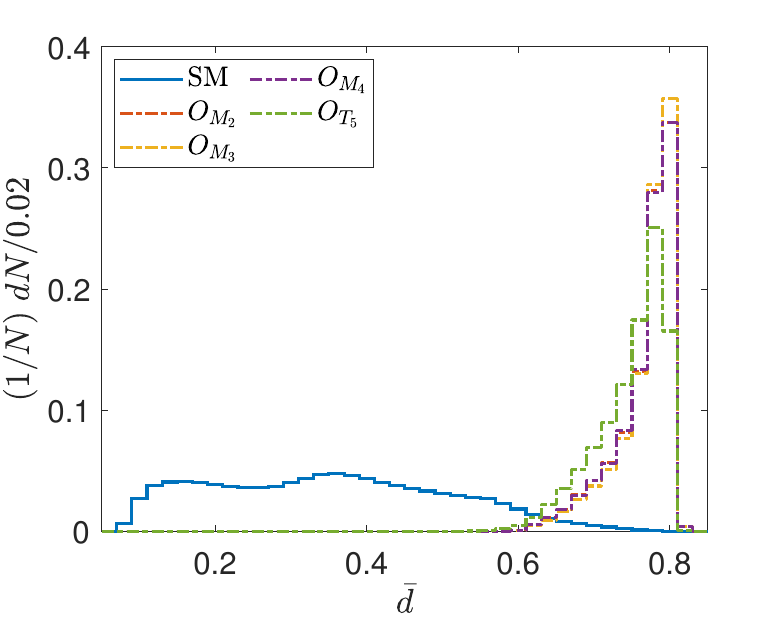}
\includegraphics[width=0.48\hsize]{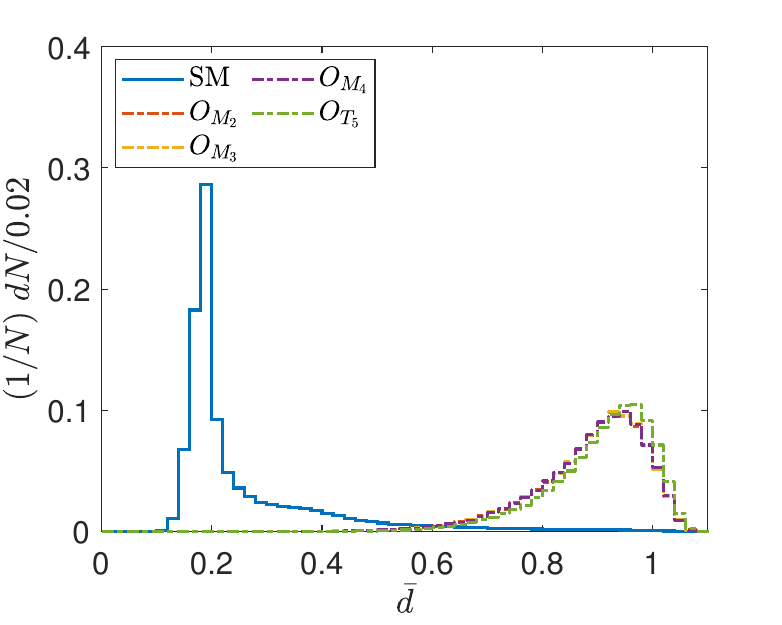}
\includegraphics[width=0.48\hsize]{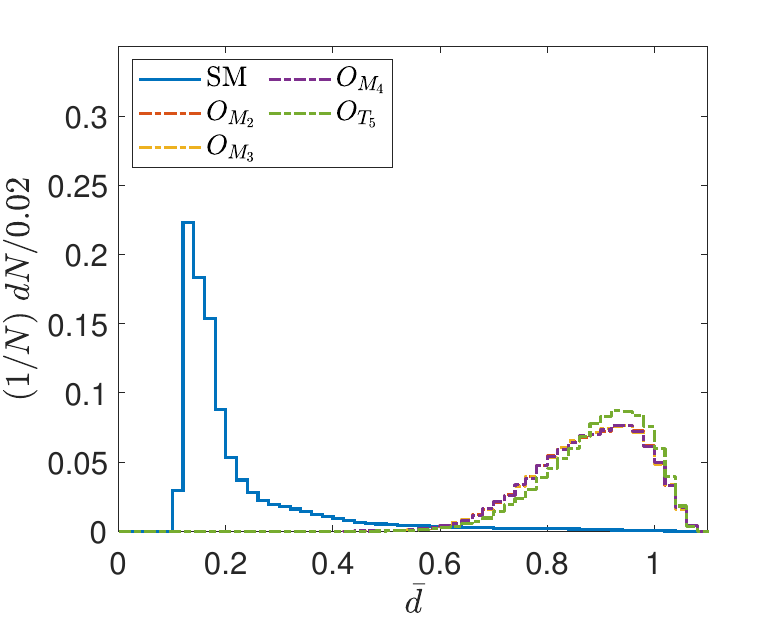}
\caption{\label{fig:rt}The normalized distribution of anomaly score $\bar{d}$ when $l=30$ for the real vector kernel~(the first row) and hardware-efficient kernel~(the second row), at $10\;{\rm TeV}$~(the first column) and $14\;{\rm TeV}$~(the second column) for the SM and $O_{M_{2,3,4}}$ and $O_{T_{5}}$ induced contribution.  }
\end{center}
\end{figure}

Another tunable parameter is $l$. 
In general, an increase in the value of $l$ leads to an improvement in the sampling of the background event distribution, as evidenced in reference~\cite{Zhang:2023yfg}. 
However, this comes at the cost of greater computational resources being required. 
Accordingly, an appropriate value of $l$ is selected to achieve an optimal balance between accuracy and computational efficiency.
Fig.~\ref{fig:l} shows the normalized distributions of anomaly scores for the SM and NP events under different $l$ values in the case of complex vector kernel.
It is evident that the anomaly score distributions for the SM background and NP events are more distinct with larger $l$. 
The value of $l$ is set to $30$ for all kernels in this study. 
The resulting normalized distributions of the real vector kernel and the hardware-efficient kernel are shown in Fig.~\ref{fig:rt}.
From the Figs.~\ref{fig:l} and \ref{fig:rt}, it can be seen that the distributions of $\bar{d}$ for the SM background and the NP signals are different, the $\bar{d}$ of the SM events are generally less than those of the NP events.
From Fig.~\ref{fig:rt}, it can also be observed that while hardware-efficient kernel shows good discrimination between the SM and NP, there is a small tail for the SM events residuals within the NP region.

\section{\label{sec4}expected coefficient constraints on the coefficients}

\begin{table}[htbp]
\centering
\begin{tabular}{c|c|c|c} 
\hline
\multicolumn{4}{c}{$\sqrt{s}=10\;{\rm TeV}$}\\
\hline
Operator &$N$&$N_{\gamma}\geq 2$ &$\epsilon_{\gamma}$ \\
\hline
 SM&$10^{6}$& $830584$ &  $83.058\%$\\
\hline
 $O_{M_{2}}$&$10^{5}$& $85261$ &  $85.261\%$\\ 
\hline
 $O_{M_{3}}$&$10^{5}$& $85527$ &  $85.527\%$\\ 
\hline
 $O_{M_{4}}$&$10^{5}$& $85481$ &  $85.481\%$\\
\hline
 $O_{T_{5}}$&$10^{5}$& $85414$ &  $85.414\%$\\
\hline
\multicolumn{4}{c}{$\sqrt{s}=14\;{\rm TeV}$}\\
\hline
Operator &$N$&$N_{\gamma}\geq 2$ &$\epsilon_{\gamma}$ \\
\hline
 SM&$2.0\times 10^{6}$& $1661890$ &  $83.095\%$\\
\hline
 $O_{M_{2}}$&$10^{5}$& $85101$ &  $85.101\%$\\
\hline
 $O_{M_{3}}$&$10^{5}$& $85591$ &  $85.591\%$\\
\hline
$O_{M_{4}}$&$10^{5}$& $85347$ &  $85.347\%$\\
\hline
 $O_{T_{5}}$&$10^{5}$& $85216$ &  $85.216\%$\\
\hline
\end{tabular}
\caption{Contributions of SM and aQGCs after $N_{\gamma}$ cut at different energies. 
$N$ and $N_{\gamma}$ represent the number of events before and after $N_{\gamma}$ cut, respectively. 
$\epsilon_{\gamma}$ is shown in the last row.}
\label{table:combined}
\end{table}

Ignoring the interference between the SM and the aQGCs, the cross-section
after cut can be expressed as,
\begin{equation}
\begin{split}
&\sigma=\epsilon_{\gamma}^{\rm SM}\epsilon_{\alpha}^{\rm SM}\sigma_{\rm SM}+ \epsilon_{\gamma}^{\rm NP}\epsilon_{\alpha}^{\rm NP}\frac{f^2}{\tilde{f}^2} \sigma_{\rm NP}
\end{split}
\label{eq.crossection}
\end{equation}
where $\sigma_{\rm SM}$ and $\sigma_{\rm NP}$ are cross-sections of the SM and NP contributions, respectively. 
The NP contribution is the one when $f_{X}=\tilde{f}_X$, where $f_X$ is the operator coefficient, and $\tilde{f}_X$ is the upper bounds of partial wave unitarity bounds listed in Table~\ref{table:unitaritybounds}. 
$\epsilon_{\gamma}^{\rm SM}$ and $\epsilon_{\gamma}^{\rm NP}$ are the cut efficiencies of the $N_{\gamma}$ cut, $\epsilon_{\alpha}^{\rm SM}$ and $\epsilon_{\alpha}^{\rm NP}$ are the cut efficiencies of the QKKM event selection strategy.
Numerical results of $\sigma_{\rm SM}$ and $\sigma_{\rm NP}$ are listed in Table~\ref{table:interference}. 
$\epsilon_{\gamma}^{\rm SM}$, $\epsilon_{\gamma}^{\rm NP}$ are listed in Table~\ref{table:combined}.

The expected coefficient constraints after cuts can be estimated by the signal significance defined as,
\begin{equation}
\begin{split}
&\mathcal{S}_{stat}=\sqrt{2 \left[(N_{\rm bg}+N_{s}) \ln (1+N_{s}/N_{\rm bg})-N_{s}\right]},
\end{split}
\label{eq.ss}
\end{equation}
where $N_{s,bg}$ are the event numbers of the signal and background, $N_{s}=(\epsilon_{\gamma}^{\rm NP}\epsilon_{\alpha}^{\rm NP}\sigma_{\rm NP}{f^2}/{f_{i}^2} )L$ and $N_{bg}=(\epsilon_{\gamma}^{\rm SM}\epsilon_{\alpha}^{\rm SM}\sigma_{\rm SM})L$, and $L$ is the luminosity.
The integrated luminosities in both ``conservative" and ``optimistic" cases~\cite{Black:2022cth,Accettura:2023ked} are considered.

\begin{figure}[htbp]
\begin{center}
\includegraphics[width=0.6\hsize]{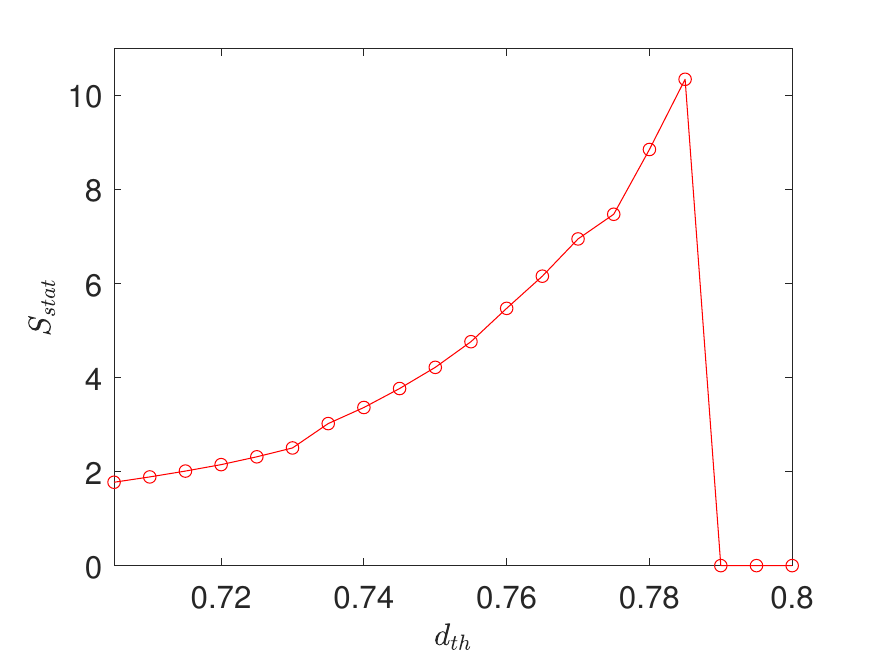}
\caption{\label{fig:sd}$S_{stat}$ as a function of $d_{th}$ for the real vector kernel in the case of $O_{M_2}$ at $\sqrt{s}=10\;\rm TeV$.
}
\end{center}
\end{figure}

\begin{table}[htbp]
\centering
\begin{tabular}{c|c|c} 
\hline
kernel&$10\;\rm TeV$&$14\;\rm TeV$\\
\hline
real vector kernel&$0.79$&$0.79$\\
\hline
complex vector kernel&$0.74$&$0.74$\\
\hline
hardware-efficient kernel&$0.75$&$0.73$\\
\hline
classical kernel&$26$&$30$\\
\hline
\end{tabular}
\caption{Values of thresholds $d_{th}$ for real vector, complex vector, hardware efficient and classical kernel for operators $O_{M_{2,3,4}}$ and $O_{T_{5}}$ at $\sqrt{s}=10\;{\rm TeV}$ and $14\;{\rm TeV}$.}
\label{table:dth}
\end{table}

To maximize signal significance, an appropriate threshold value $d_{th}$ is selected. 
Taking the real vector kernel operator $O_{M_2}$ at $10\; \rm TeV$ as an example. 
As shown in Fig.~\ref{fig:sd}, $S_{stat}$ varies with $d_{th}$ within the given range. 
The corresponding $d_{th}$ value is chosen as the final threshold when $S_{stat}$ reaches its maximum. 
The shape of the function $S_{stat}(d_{th})$ in other scenarios is similar to that shown in Fig.~\ref{fig:sd}.
The results of $d_{th}$ for the real vector kernel, the complex vector kernel, and the hardware-efficient kernel are listed in Table~\ref{table:dth}. 
To compare the results of quantum and classical algorithm, the classical kernel is included.
The expected coefficient constraints are calculated using the classical kernel from the \verb"scikit-learn"~\cite{Pedregosa:2011ork} package, following the same steps as in Ref.~\cite{Zhang:2023yfg}.
The $d_{th}$ for the classical kernel is also listed in Table~\ref{table:dth}.
In quantum computing, since the kernel is computed using inner products, the value of $d_{th}$ always lies between $0$ and $1$, which is not the case for the classical k-means where the definition of $d$ is different which is the Euclidean distance~\cite{Zhang:2023yfg}.

\begin{table}[htbp]
\centering
\begin{tabular}{c|c|c} 
\hline
\multicolumn{3}{c}{real vector kernel}\\
\hline
$\sqrt{s}$&$10\;{\rm TeV}$&$14\;{\rm TeV}$\\
\hline
Operator&$\epsilon_{\alpha}$($a>0.79$)&$\epsilon_{\alpha}$($a>0.79$)\\
\hline
SM&$0.000482\%$& $0.0000602\%$\\
\hline
$O_{M_{2}}$&$20.155\%$&$34.172\%$\\  
\hline
 $O_{M_{3}}$&$21.773\%$&$36.135\%$\\ 
\hline
$O_{M_{4}}$&$19.954\%$&$34.151\%$\\  
\hline
$O_{T_{5}}$&$9.713\%$&$16.620\%$\\
\hline
\multicolumn{3}{c}{complex vector kernel}\\
\hline
$\sqrt{s}$&$10\;{\rm TeV}$&$14\;{\rm TeV}$\\
\hline
Operator&$\epsilon_{\alpha}$($a>0.74$)&$\epsilon_{\alpha}$($a>0.74$)\\
\hline
SM&$0.000120\%$& $0.000602\%$\\
\hline
$O_{M_{2}}$&$4.987\%$&$10.708\%$\\  
\hline
 $O_{M_{3}}$&$5.444\%$&$11.416\%$\\ 
\hline
$O_{M_{4}}$&$4.828\%$&$10.689\%$\\  
\hline
$O_{T_{5}}$&$2.750\%$&$6.491\%$\\
\hline
\multicolumn{3}{c}{hardware-efficient kernel}\\
\hline
$\sqrt{s}$&$10\;{\rm TeV}$&$14\;{\rm TeV}$\\
\hline
Operator&$\epsilon_{\alpha}$($a>0.75$)&$\epsilon_{\alpha}$($a>0.73$)\\
\hline
SM&$0.954\%$& $0.965\%$\\
\hline
$O_{M_{2}}$&$71.063\%$&$68.378\%$\\  
\hline
 $O_{M_{3}}$&$71.134\%$&$69.676\%$\\ 
\hline
$O_{M_{4}}$&$71.264\%$&$69.823\%$\\  
\hline
$O_{T_{5}}$&$77.247\%$&$77.030\%$\\
\hline
\multicolumn{3}{c}{classical kernel}\\
\hline
$\sqrt{s}$&$10\;{\rm TeV}$&$14\;{\rm TeV}$\\
\hline
Operator&$\epsilon_{\alpha}$($a>0.75$)&$\epsilon_{\alpha}$($a>0.73$)\\
\hline
SM&$0.321\%$& $0.223\%$\\
\hline
$O_{M_{2}}$&$56.449\%$&$60.673\%$\\  
\hline
 $O_{M_{3}}$&$57.896\%$&$62.163\%$\\ 
\hline
$O_{M_{4}}$&$56.453\%$&$60.934\%$\\  
\hline
$O_{T_{5}}$&$61.635\%$&$65.577\%$\\
\hline
\end{tabular}
\caption{For the different kernels, contributions of SM and aQGCs after QKKM cut at different energies. 
The $a$ is the anomaly score and the efficiency of the event selection strategy $\epsilon_{\alpha}$ is shown in the last row.}
\label{table:cutrk}
\end{table}

The $\epsilon_{\alpha}^{\rm SM}$ and $\epsilon_{\alpha}^{\rm NP}$ are defined as the cut efficiencies of QKKM event selection strategy.
The cut efficiencies of the four different kernels when the $d_{th}$ are chosen as the ones listed in Table~\ref{table:dth} are listed in Table~\ref{table:cutrk}.
As can be seen from Table~\ref{table:cutrk}, for the complex vector kernel, a relatively strict $d_{th}$ is taken, which is due to the fact that the background can be suppressed to a very low level. 
For hardware-efficient kernel, a relatively loose $d_{th}$ is taken due to the fact that the tail of the background events in the NP region. 
All the $d_{th}$s are chosen according to $\mathcal{S}_{stat}$.

\begin{table}[htbp]
\centering
\begin{tabular}{c|c|c|c|c} 
\hline
  &$S_{stat}$&10\;{\rm TeV} &14\;{\rm TeV}&14\;{\rm TeV} \\
  & &$10\;{\rm ab}^{-1}$&$10\;{\rm ab}^{-1}$&$20\;{\rm ab}^{-1}$  \\
   & &$(\rm TeV^{-4})$&$(\rm TeV^{-4})$&$(\rm TeV^{-4})$  \\
\hline
  & $2$& $<5.39\times 10^{-3}$&$<1.89\times 10^{-3}$&$<1.56\times 10^{-3}$\\
$\frac{f_{M_{2}}}{\Lambda^4}$& 3 & $<6.91\times 10^{-3}$&$<2.38\times 10^{-3}$& $<1.96\times 10^{-3}$  \\
  & $5$ &$<9.63\times 10^{-3}$&$<3.22\times 10^{-3}$&$<2.62\times 10^{-3}$\\
\hline
 & $2$ &$<1.93\times 10^{-3}$&$<6.85\times 10^{-4}$&$<5.67\times 10^{-4}$ \\
$\frac{f_{M_{3}}}{\Lambda^4}$& $3$ &$<2.47\times 10^{-3}$&$<8.62\times 10^{-4}$&$<7.08\times 10^{-4}$\\
  & $5$ &$<3.45\times 10^{-3}$&$<1.17\times 10^{-3}$&$<9.48\times 10^{-4}$\\
\hline
 &$ 2$ &$<1.98\times 10^{-3}$&$<6.85\times 10^{-3}$&$<5.66\times 10^{-3}$\\
$\frac{f_{M_{4}}}{\Lambda^4}$& $3$ &$<2.54\times 10^{-3}$&$<8.61\times 10^{-3}$&$<3.54\times 10^{-3}$\\
 & $5$ &$<4.00\times 10^{-3}$&$<1.16\times 10^{-2}$&$<9.48\times 10^{-3}$\\
\hline
& $2$ &$<5.16\times 10^{-4}$&$<1.62\times 10^{-3}$&$<1.34\times 10^{-3}$\\
$\frac{f_{T_{5}}}{\Lambda^4}$& $3$ &$<6.62\times 10^{-4}$&$<2.04\times 10^{-3}$&$<1.68\times 10^{-3}$\\
  & $5$&$<9.22\times 10^{-4}$&$<2.76\times 10^{-3}$&$<2.24\times 10^{-3}$ \\
\hline
\end{tabular}
\caption{Projected sensitivity the coefficients of the 
$O_{M_{2,3,4}}$ and $O_{T_{5}}$ operators at muon colliders in the `conservative' and `optimistic' cases when the kernel function is complex vector kernel.}
\label{table:conservative}
\end{table}

\begin{table}[htbp]
\centering
\begin{tabular}{c|c|c|c|c} 
\hline
  &$S_{stat}$&10\;{\rm TeV} &14\;{\rm TeV}&14\;{\rm TeV} \\
  & &$10\;{\rm ab}^{-1}$&$10\;{\rm ab}^{-1}$&$20\;{\rm ab}^{-1}$  \\
   & &$(\rm TeV^{-4})$&$(\rm TeV^{-4})$&$(\rm TeV^{-4})$  \\
\hline
  & 2& $<3.58\times 10^{-3}$&$<6.58\times 10^{-4}$&$<5.33\times 10^{-4}$\\
$\frac{f_{M_{2}}}{\Lambda^4}$& 3 & $<4.52\times 10^{-3}$&$<8.53\times 10^{-4}$& $<6.83\times 10^{-4}$  \\
  & 5 &$<6.14\times 10^{-3}$&$<1.20\times 10^{-3}$&$<9.51\times 10^{-4}$\\
\hline
 & 2 &$<1.29\times 10^{-3}$&$<2.40\times 10^{-4}$&$<1.94\times 10^{-4}$ \\
$\frac{f_{M_{3}}}{\Lambda^4}$& 3 &$<1.63\times 10^{-3}$&$<3.10\times 10^{-4}$&$<2.49\times 10^{-4}$\\
  & 5 &$<2.21\times 10^{-3}$&$<4.38\times 10^{-4}$&$<3.46\times 10^{-4}$\\
\hline
 & 2 &$<1.30\times 10^{-3}$&$<2.38\times 10^{-3}$&$<1.93\times 10^{-3}$\\
$\frac{f_{M_{4}}}{\Lambda^4}$& 3 &$<1.64\times 10^{-3}$&$<3.09\times 10^{-3}$&$<2.47\times 10^{-3}$\\
 & 5 &$<2.23\times 10^{-3}$&$<4.36\times 10^{-3}$&$<3.44\times 10^{-3}$\\
\hline
& 2 &$<3.67\times 10^{-4}$&$<6.30\times 10^{-4}$&$<5.09\times 10^{-4}$\\
$\frac{f_{T_{5}}}{\Lambda^4}$& 3 &$<4.63\times 10^{-4}$&$<8.16\times 10^{-4}$&$<6.53\times 10^{-4}$\\
  & 5&$<6.29\times 10^{-4}$&$<1.15\times 10^{-3}$&$<9.10\times 10^{-4}$ \\
\hline
\end{tabular}
\caption{Same as Table~\ref{table:conservative}, but for the real vector kernel.}
\label{table:rconservative}
\end{table}

\begin{table}[htbp]
\centering
\begin{tabular}{c|c|c|c|c} 
\hline
  &$S_{stat}$&10\;{\rm TeV} &14\;{\rm TeV}&14\;{\rm TeV} \\
  & &$10\;{\rm ab}^{-1}$&$10\;{\rm ab}^{-1}$&$20\;{\rm ab}^{-1}$  \\
   & &$(\rm TeV^{-4})$&$(\rm TeV^{-4})$&$(\rm TeV^{-4})$  \\
\hline
  & 2&$<1.19\times 10^{-2}$& $<4.42\times 10^{-3}$&$<3.72\times 10^{-3}$\\
$\frac{f_{M_{2}}}{\Lambda^4}$& 3 &$<1.46\times 10^{-2}$& $<5.42\times 10^{-3}$& $<4.56\times 10^{-3}$  \\
  & 5 &$<1.89\times 10^{-2}$&$<7.01\times 10^{-3}$&$<5.89\times 10^{-3}$\\
\hline
 & 2 &$<4.46\times 10^{-3}$&$<1.65\times 10^{-3}$&$<1.39\times 10^{-3}$ \\
$\frac{f_{M_{3}}}{\Lambda^4}$& 3 &$<5.46\times 10^{-3}$&$<2.02\times 10^{-3}$&$<1.70\times 10^{-3}$\\
  & 5 &$<7.07\times 10^{-3}$&$<2.62\times 10^{-3}$&$<2.20\times 10^{-3}$\\
\hline
 & 2 &$<4.30\times 10^{-3}$&$<1.59\times 10^{-2}$&$<1.34\times 10^{-2}$\\
$\frac{f_{M_{4}}}{\Lambda^4}$& 3 &$<5.27\times 10^{-3}$&$<1.96\times 10^{-2}$&$<1.64\times 10^{-2}$\\
 & 5 &$<6.82\times 10^{-3}$&$<2.53\times 10^{-2}$&$<2.13\times 10^{-2}$\\
\hline
& 2 &$<8.11\times 10^{-4}$&$<2.80\times 10^{-3}$&$<2.36\times 10^{-3}$\\
$\frac{f_{T_{5}}}{\Lambda^4}$& 3 &$<9.95\times 10^{-4}$&$<3.44\times 10^{-3}$&$<2.89\times 10^{-3}$\\
  & 5&$<1.29\times 10^{-3}$&$<4.44\times 10^{-3}$&$<3.73\times 10^{-3}$ \\
\hline
\end{tabular}
\caption{Same as Table~\ref{table:conservative}, but for the hardware-efficient kernel.}
\label{table:cconservative}
\end{table}

\begin{table}[htbp]
\centering
\begin{tabular}{c|c|c|c|c} 
\hline
  &$S_{stat}$&10\;{\rm TeV} &14\;{\rm TeV}&14\;{\rm TeV} \\
  & &$10\;{\rm ab}^{-1}$&$10\;{\rm ab}^{-1}$&$20\;{\rm ab}^{-1}$  \\
   & &$(\rm TeV^{-4})$&$(\rm TeV^{-4})$&$(\rm TeV^{-4})$  \\
\hline
  & 2& $<5.77\times 10^{-3}$&$<1.86\times 10^{-3}$&$<1.56\times 10^{-3}$\\
$\frac{f_{M_{2}}}{\Lambda^4}$& 3 & $<7.11\times 10^{-3}$&$<2.29\times 10^{-3}$& $<1.92\times 10^{-3}$  \\
  & 5 &$<9.27\times 10^{-3}$&$<2.99\times 10^{-3}$&$<2.50\times 10^{-3}$\\
\hline
 & 2 &$<2.13\times 10^{-3}$&$<6.88\times 10^{-4}$&$<5.77\times 10^{-4}$ \\
$\frac{f_{M_{3}}}{\Lambda^4}$& 3 &$<2.62\times 10^{-3}$&$<8.48\times 10^{-4}$&$<7.10\times 10^{-4}$\\
  & 5 &$<3.41\times 10^{-3}$&$<1.11\times 10^{-3}$&$<9.23\times 10^{-4}$\\
\hline
 & 2 &$<2.09\times 10^{-3}$&$<6.72\times 10^{-3}$&$<5.63\times 10^{-3}$\\
$\frac{f_{M_{4}}}{\Lambda^4}$& 3 &$<2.57\times 10^{-3}$&$<8.28\times 10^{-3}$&$<6.93\times 10^{-3}$\\
 & 5 &$<3.34\times 10^{-3}$&$<1.08\times 10^{-2}$&$<9.01\times 10^{-3}$\\
\hline
& 2 &$<3.92\times 10^{-4}$&$<1.20\times 10^{-3}$&$<1.00\times 10^{-3}$\\
$\frac{f_{T_{5}}}{\Lambda^4}$& 3 &$<4.83\times 10^{-4}$&$<1.47\times 10^{-3}$&$<1.23\times 10^{-3}$\\
  & 5&$<6.29\times 10^{-4}$&$<1.92\times 10^{-3}$&$<1.60\times 10^{-3}$ \\
\hline
\end{tabular}
\caption{The same as Table~\ref{table:conservative}, but using classical k-means.}
\label{table:kconservative}
\end{table}

\begin{table}[hbtp]
\centering
\begin{tabular}{c|c|c|c|c}
\hline
coefficient &$f_{M_{2}}/\Lambda^4$ &$f_{M_{3}}/\Lambda^4$ &$f_{M_{4}}/\Lambda^4$ &$f_{T_{5}}/\Lambda^4$ \\
\hline
constraint&$[-2.8, 2.8]$&$[-4.4, 4.4]$&$[-5.0, 5.0]$&$[-0.5, 0.5]$ \\
\hline
\end{tabular}
\caption{The constraints on the $O_{M_{i}}$ and $O_{T_{i}}$ coefficients (${\rm TeV}^{-4}$) obtained at $95\%$ C.L at the LHC~\cite{CMS:2019qfk,CMS:2020ypo}.}
\label{table:constraints}
\end{table}

When $d_{th}$ is chosen, the expected coefficient constraints can be obtained by using signal significance.
The results of the expected coefficient constraints in the case of complex vector kernel, real vector kernel, hard-efficient kernel, and the classical kernel are shown in Tables~\ref{table:conservative}, \ref{table:rconservative}, \ref{table:cconservative}, and \ref{table:kconservative}, respectively.
It can be seen that the muon collider with $\sqrt{s} \geq 10\;\rm TeV$ has tighter constraints than the ones at the LHC~\cite{CMS:2019qfk,CMS:2020ypo} in Table~\ref{table:constraints}.
We speculate that this is due to the fact that, compared to the classical case, the Hilbert space in which the data resides is of higher dimensionality and thus the data is better separable.
The sensitivities of the muon colliders to the aQGCs are competitive with future hadron colliders and even better at the same c.m. energy.
The muon colliders are suitable to study the aQGCs because of the high energies and luminosities as well as having a cleaner experimental environment than hadron colliders.

\begin{figure}[htbp]
\begin{center}
\includegraphics[width=0.8\hsize]{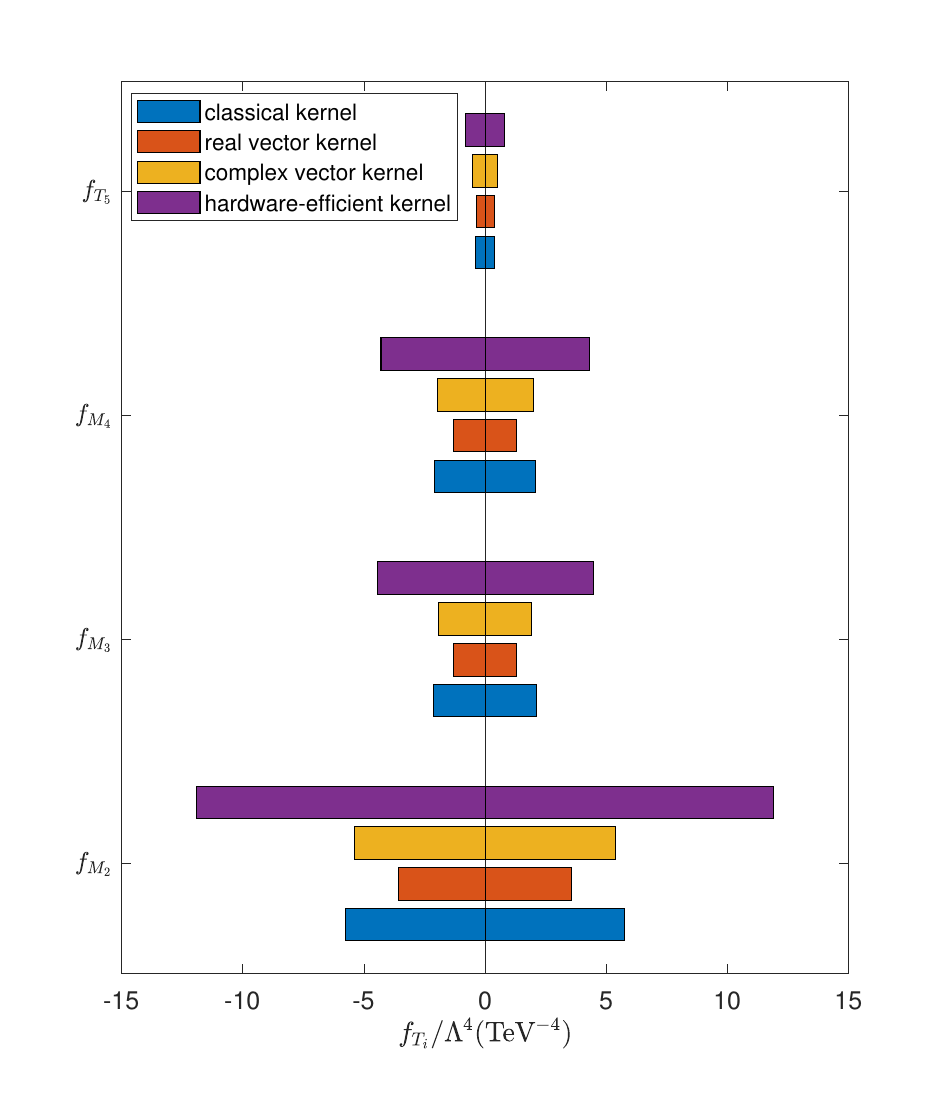}
\caption{\label{fig:compare}Comparison of the expected coefficient constraints obtained when the kernel function is classical kernel,real vector kernel and complex vector kernel, as well as hardware-efficient kernel, at $\sqrt{s}=10\;\rm TeV$, $S_{stat}=2$.}
\end{center}
\end{figure}

For comparison, we take $\sqrt{s} = 10\;\rm{TeV}$ and $S_{stat}=2$ as an example, as can be seen from the Fig.~\ref{fig:compare}, the expected coefficient constraints are tightest for coefficients $f_{M_{2,3,4}}$ and $f_{T_5}$ when the kernel is real vector kernel. 
This shows that instead of affecting the performance of k-means, the quantum kernel function outperforms a classical kernel.
While the SM and NP signals are most effectively distinguished when utilizing the hardware-efficient kernel, the fact that the SM leaves small residuals in the NP signals results in the least stringent constraints.

After all, it can be conclude that the QKKM algorithm is an effective tool to search for the NP signals.
The real vector kernel works better than the classical kernel, not to mention the potential that the QKKM can cope with future developments in quantum computing for example when the MC data is generated by a quantum computer.
Note that for all the quantum kernels, the matrix elements can be calculated using swap test, therefore can be accelerated by multi-state swap test~\cite{Liu:2022jsp,Fanizza:2020qjq}.

At this stage, the effect of noise in quantum computing is unavoidable.
In comparison with the Ref.~\cite{Zhang:2023ykh}, the quantum computer has the same task of computing the kernel matrix, two of the quantum kernel functions~(real and complex vector kernels) used are the same, the dimensions of the vectors dealt with are of the same order of magnitude, and thus the number of qubits used is the same, and the size of the datasets are also of the same order of magnitude. 
Therefore, we directly borrow the results from the Ref.~\cite{Zhang:2023ykh} to estimate the effect of noise.
According to Ref.~\cite{Zhang:2023ykh}, the noise-induced relative errors when using the real and complex vector kernels can be estimated to be about $6.25\%$. 
In addition, the error from noise induced by the hardware-efficient kernel is expected to be even smaller, because the quantum circuit of the hardware-efficient kernel does not contain CNOT gates.
Therefore, the above error value is the upper limit of the noise-induced error of the hardware-efficient kernel.

\section{\label{sec5}Summary}

The search for new physics~(NP) signals requires the processing of large volumes of data, and quantum computing has the potential to accelerate these computations in the future.
This paper focuses on the $\mu^+\mu^-\rightarrow \nu\bar{\nu}\gamma\gamma$ process at a muon collider, a process highly sensitive to dimension-8 operators involved in anomalous quartic gauge couplings (aQGCs). 
We use kernel K-means AD to search for the signals of the NP.
In this paper, three different types of quantum kernels and a classical kernel are used.

The results indicate that this process is indeed highly sensitive to the aQGCs. 
The kernel K-means AD algorithm, utilizing the three distinct quantum kernels~(when a quantum kernel is used, it is QKKM), as well as the classical kernel-based algorithm, proved feasible for NP signal searches.
Compared to the LHC, the muon collider offers more stringent coefficient constraints.
Among the four kernels, the real vector kernel demonstrated the best performance. 
Therefore, it is suggested that the QKKM is well suited for the phenomenological study of the NP, especially when progress in quantum computing are anticipated.

\begin{acknowledgements}
This work was supported in part by the National Natural Science Foundation of China under Grants No.~12147214, the Natural Science Foundation of the Liaoning Scientific Committee Nos.~LJKZ0978 and LJKMZ20221431.
\end{acknowledgements}

\appendix

\section{\label{secap}Contributions of tri-boson and VBS processes for \texorpdfstring{$O_{M}$}{OM} operators}

Contributions of tri-boson and VBS processes for $O_{T_5}$ operator is established in Ref.~\cite{Yang:2020rjt}.
For $O_{M}$ operators, using effective vector boson approximation~\cite{Kane:1984bb,Boos:1997gw,Ruiz:2021tdt}, at the leading order of $\mathcal{O}(M_Z^2/s)$,
\begin{equation}
\begin{split}
&\sigma_{NP}^{VBS}=\frac{e^8 s^3 v^4 }{36238786560 \pi ^5 M_W^4 s_W^8}\left[15 \left(4 c_W^2 (4 f_{M_2}-f_{M_3})\right.\right.\\
&\left.\left.+4 c_W s_W (2 f_{M_4}+f_{M_5})+s_W^2 (8 f_{M_0}-f_{17})\right)^2\right.\\
&\left.+2 \left(-4 c_W^2 f_{M_3}+4 c_Ws_W f_{M_5} -s_W^2 f_{17}\right)^2\right],
\end{split}  
\end{equation}
with $f_{17}=2f_{M_1}-f_{M_7}$.
For the tri-boson case, at the leading order of $\mathcal{O}(M_Z^2/s)$,
\begin{equation}
\begin{split}
&\sigma_{NP}^{t.-b.}={\rm Br}(Z\to \nu\bar{\nu})\times \frac{e^2 M_Z^2 s^2 \left(c_W^4-2 c_W^2 s_W^2+5 s_W^4\right)}{283115520 \pi ^3 c_W^2 s_W^2} \\
&\times \left\{16 c_W^4 \left(48 f_{M_2}^2-24 f_{M_2} f_{M_3}+19 f_{M_3}^2\right)\right.\\
&\left.-32 c_W^3 s_W \left[12 f_{M_2} (2 f_{M_4}+f_{M_5})-f_{M_3} (6 f_{M_4}+19 f_{M_5})\right]\right.\\
&\left.+8 c_W^2 s_W^2 \left[24 f_{M_0} (4 f_{M_2}-f_{M_3})+24 f_{M_4}^2+24f_{M_4} f_{M_5}\right.\right.\\
&\left.\left.+38 f_{M_5}^2-f_{17} (12 f_{M_2}-19 f_{M_3})\right]\right.\\
&\left.-8 c_W s_W^3 \left[24 f_{M_0} (2 f_{M_4}+f_{M_5})-f_{17} (6 f_{M_4}+19 f_{M_5})\right]\right.\\
&\left.+s_W^4 \left[192 f_{M_0}^2-48 f_{M_0} f_{17}+19 f_{17}^2\right]\right\}.
\end{split}  
\end{equation}

\bibliography{QKK}

\begin{thebibliography}{100}
\expandafter\ifx\csname url\endcsname\relax
  \def\url#1{\texttt{#1}}\fi
\expandafter\ifx\csname urlprefix\endcsname\relax\def\urlprefix{URL }\fi
\expandafter\ifx\csname href\endcsname\relax
  \def\href#1#2{#2} \def\path#1{#1}\fi

\bibitem{Arute:2019zxq}
F.~Arute, et~al., {Quantum supremacy using a programmable superconducting
  processor}, Nature 574~(7779) (2019) 505--510.
\newblock \href {http://arxiv.org/abs/1910.11333} {\path{arXiv:1910.11333}},
  \href {https://doi.org/10.1038/s41586-019-1666-5}
  {\path{doi:10.1038/s41586-019-1666-5}}.

\bibitem{Ellis:2012zz}
J.~Ellis, {Outstanding questions: Physics beyond the Standard Model}, Phil.
  Trans. Roy. Soc. Lond. A 370 (2012) 818--830.
\newblock \href {https://doi.org/10.1098/rsta.2011.0452}
  {\path{doi:10.1098/rsta.2011.0452}}.

\bibitem{Preskill:2018jim}
J.~Preskill, {Quantum Computing in the NISQ era and beyond}, Quantum 2 (2018)
  79.
\newblock \href {http://arxiv.org/abs/1801.00862} {\path{arXiv:1801.00862}},
  \href {https://doi.org/10.22331/q-2018-08-06-79}
  {\path{doi:10.22331/q-2018-08-06-79}}.

\bibitem{CG2023}
C.~Gill,
  \href{https://cra.org/ccc/events/5-year-update-to-the-next-steps-in-quantum-computing-workshop/}{5
  year update to the next steps in quantum computing workshop}, Computing
  Community Consortium (2023).
\newline\urlprefix\url{https://cra.org/ccc/events/5-year-update-to-the-next-steps-in-quantum-computing-workshop/}

\bibitem{Zhu:2024own}
Y.~Zhu, W.~Zhuang, C.~Qian, Y.~Ma, D.~E. Liu, M.~Ruan, C.~Zhou, {A Novel
  Quantum Realization of Jet Clustering in High-Energy Physics Experiments},
  2024.
\newblock \href {http://arxiv.org/abs/2407.09056} {\path{arXiv:2407.09056}}.

\bibitem{Carena:2022kpg}
M.~Carena, H.~Lamm, Y.-Y. Li, W.~Liu, {Improved Hamiltonians for Quantum
  Simulations of Gauge Theories}, Phys. Rev. Lett. 129~(5) (2022) 051601.
\newblock \href {http://arxiv.org/abs/2203.02823} {\path{arXiv:2203.02823}},
  \href {https://doi.org/10.1103/PhysRevLett.129.051601}
  {\path{doi:10.1103/PhysRevLett.129.051601}}.

\bibitem{Bauer:2022hpo}
C.~W. Bauer, et~al., {Quantum Simulation for High-Energy Physics}, PRX Quantum
  4~(2) (2023) 027001.
\newblock \href {http://arxiv.org/abs/2204.03381} {\path{arXiv:2204.03381}},
  \href {https://doi.org/10.1103/PRXQuantum.4.027001}
  {\path{doi:10.1103/PRXQuantum.4.027001}}.

\bibitem{Roggero:2018hrn}
A.~Roggero, J.~Carlson, {Dynamic linear response quantum algorithm}, Phys. Rev.
  C 100~(3) (2019) 034610.
\newblock \href {http://arxiv.org/abs/1804.01505} {\path{arXiv:1804.01505}},
  \href {https://doi.org/10.1103/PhysRevC.100.034610}
  {\path{doi:10.1103/PhysRevC.100.034610}}.

\bibitem{Roggero:2019myu}
A.~Roggero, A.~C.~Y. Li, J.~Carlson, R.~Gupta, G.~N. Perdue, {Quantum Computing
  for Neutrino-Nucleus Scattering}, Phys. Rev. D 101~(7) (2020) 074038.
\newblock \href {http://arxiv.org/abs/1911.06368} {\path{arXiv:1911.06368}},
  \href {https://doi.org/10.1103/PhysRevD.101.074038}
  {\path{doi:10.1103/PhysRevD.101.074038}}.

\bibitem{Gustafson:2022xdt}
E.~J. Gustafson, H.~Lamm, F.~Lovelace, D.~Musk, {Primitive quantum gates for an
  SU(2) discrete subgroup: Binary tetrahedral}, Phys. Rev. D 106~(11) (2022)
  114501.
\newblock \href {http://arxiv.org/abs/2208.12309} {\path{arXiv:2208.12309}},
  \href {https://doi.org/10.1103/PhysRevD.106.114501}
  {\path{doi:10.1103/PhysRevD.106.114501}}.

\bibitem{Lamm:2024jnl}
H.~Lamm, Y.-Y. Li, J.~Shu, Y.-L. Wang, B.~Xu, {Block encodings of discrete
  subgroups on a quantum computer}, Phys. Rev. D 110~(5) (2024) 054505.
\newblock \href {http://arxiv.org/abs/2405.12890} {\path{arXiv:2405.12890}},
  \href {https://doi.org/10.1103/PhysRevD.110.054505}
  {\path{doi:10.1103/PhysRevD.110.054505}}.

\bibitem{Carena:2024dzu}
M.~Carena, H.~Lamm, Y.-Y. Li, W.~Liu, {Quantum error thresholds for
  gauge-redundant digitizations of lattice field theories}, Phys. Rev. D
  110~(5) (2024) 054516.
\newblock \href {http://arxiv.org/abs/2402.16780} {\path{arXiv:2402.16780}},
  \href {https://doi.org/10.1103/PhysRevD.110.054516}
  {\path{doi:10.1103/PhysRevD.110.054516}}.

\bibitem{Atas:2021ext}
Y.~Y. Atas, J.~Zhang, R.~Lewis, A.~Jahanpour, J.~F. Haase, C.~A. Muschik,
  {SU(2) hadrons on a quantum computer via a variational approach}, Nature
  Commun. 12~(1) (2021) 6499.
\newblock \href {http://arxiv.org/abs/2102.08920} {\path{arXiv:2102.08920}},
  \href {https://doi.org/10.1038/s41467-021-26825-4}
  {\path{doi:10.1038/s41467-021-26825-4}}.

\bibitem{Li:2023vwx}
Y.-Y. Li, M.~O. Sajid, J.~Unmuth-Yockey, {Lattice holography on a quantum
  computer}, Phys. Rev. D 110~(3) (2024) 034507.
\newblock \href {http://arxiv.org/abs/2312.10544} {\path{arXiv:2312.10544}},
  \href {https://doi.org/10.1103/PhysRevD.110.034507}
  {\path{doi:10.1103/PhysRevD.110.034507}}.

\bibitem{Cui:2019sfz}
X.~Cui, Y.~Shi, J.-C. Yang, {Circuit-based digital adiabatic quantum simulation
  and pseudoquantum simulation as new approaches to lattice gauge theory}, JHEP
  08 (2020) 160.
\newblock \href {http://arxiv.org/abs/1910.08020} {\path{arXiv:1910.08020}},
  \href {https://doi.org/10.1007/JHEP08(2020)160}
  {\path{doi:10.1007/JHEP08(2020)160}}.

\bibitem{Zou:2021pvl}
Y.-T. Zou, Y.-J. Bo, J.-C. Yang, {Optimize quantum simulation using a
  force-gradient integrator}, EPL 135 (2021) 10004.
\newblock \href {http://arxiv.org/abs/2103.05876} {\path{arXiv:2103.05876}},
  \href {https://doi.org/10.1209/0295-5075/135/10004}
  {\path{doi:10.1209/0295-5075/135/10004}}.

\bibitem{Georgescu:2013oza}
I.~M. Georgescu, S.~Ashhab, F.~Nori, {Quantum Simulation}, Rev. Mod. Phys. 86
  (2014) 153.
\newblock \href {http://arxiv.org/abs/1308.6253} {\path{arXiv:1308.6253}},
  \href {https://doi.org/10.1103/RevModPhys.86.153}
  {\path{doi:10.1103/RevModPhys.86.153}}.

\bibitem{Lamm:2019uyc}
H.~Lamm, S.~Lawrence, Y.~Yamauchi, {Parton physics on a quantum computer},
  Phys. Rev. Res. 2~(1) (2020) 013272.
\newblock \href {http://arxiv.org/abs/1908.10439} {\path{arXiv:1908.10439}},
  \href {https://doi.org/10.1103/PhysRevResearch.2.013272}
  {\path{doi:10.1103/PhysRevResearch.2.013272}}.

\bibitem{Li:2021kcs}
T.~Li, X.~Guo, W.~K. Lai, X.~Liu, E.~Wang, H.~Xing, D.-B. Zhang, S.-L. Zhu,
  {Partonic collinear structure by quantum computing}, Phys. Rev. D 105~(11)
  (2022) L111502.
\newblock \href {http://arxiv.org/abs/2106.03865} {\path{arXiv:2106.03865}},
  \href {https://doi.org/10.1103/PhysRevD.105.L111502}
  {\path{doi:10.1103/PhysRevD.105.L111502}}.

\bibitem{Echevarria:2020wct}
M.~G. Echevarria, I.~L. Egusquiza, E.~Rico, G.~Schnell, {Quantum simulation of
  light-front parton correlators}, Phys. Rev. D 104~(1) (2021) 014512.
\newblock \href {http://arxiv.org/abs/2011.01275} {\path{arXiv:2011.01275}},
  \href {https://doi.org/10.1103/PhysRevD.104.014512}
  {\path{doi:10.1103/PhysRevD.104.014512}}.

\bibitem{Jordan:2011ci}
S.~P. Jordan, K.~S.~M. Lee, J.~Preskill, {Quantum Computation of Scattering in
  Scalar Quantum Field Theories}, Quant. Inf. Comput. 14 (2014) 1014--1080.
\newblock \href {http://arxiv.org/abs/1112.4833} {\path{arXiv:1112.4833}}.

\bibitem{Mueller:2019qqj}
N.~Mueller, A.~Tarasov, R.~Venugopalan, {Deeply inelastic scattering structure
  functions on a hybrid quantum computer}, Phys. Rev. D 102~(1) (2020) 016007.
\newblock \href {http://arxiv.org/abs/1908.07051} {\path{arXiv:1908.07051}},
  \href {https://doi.org/10.1103/PhysRevD.102.016007}
  {\path{doi:10.1103/PhysRevD.102.016007}}.

\bibitem{Chou:2023hcc}
A.~Chou, et~al., {Quantum Sensors for High Energy Physics}, 2023.
\newblock \href {http://arxiv.org/abs/2311.01930} {\path{arXiv:2311.01930}}.

\bibitem{Bauer:2019qxa}
C.~W. Bauer, W.~A. de~Jong, B.~Nachman, D.~Provasoli, {Quantum Algorithm for
  High Energy Physics Simulations}, Phys. Rev. Lett. 126~(6) (2021) 062001.
\newblock \href {http://arxiv.org/abs/1904.03196} {\path{arXiv:1904.03196}},
  \href {https://doi.org/10.1103/PhysRevLett.126.062001}
  {\path{doi:10.1103/PhysRevLett.126.062001}}.

\bibitem{Weinberg:1979sa}
S.~Weinberg, {Baryon and Lepton Nonconserving Processes}, Phys. Rev. Lett. 43
  (1979) 1566--1570.
\newblock \href {https://doi.org/10.1103/PhysRevLett.43.1566}
  {\path{doi:10.1103/PhysRevLett.43.1566}}.

\bibitem{Grzadkowski:2010es}
B.~Grzadkowski, M.~Iskrzynski, M.~Misiak, J.~Rosiek, {Dimension-Six Terms in
  the Standard Model Lagrangian}, JHEP 10 (2010) 085.
\newblock \href {http://arxiv.org/abs/1008.4884} {\path{arXiv:1008.4884}},
  \href {https://doi.org/10.1007/JHEP10(2010)085}
  {\path{doi:10.1007/JHEP10(2010)085}}.

\bibitem{Brivio:2017vri}
I.~Brivio, M.~Trott, {The Standard Model as an Effective Field Theory}, Phys.
  Rept. 793 (2019) 1--98.
\newblock \href {http://arxiv.org/abs/1706.08945} {\path{arXiv:1706.08945}},
  \href {https://doi.org/10.1016/j.physrep.2018.11.002}
  {\path{doi:10.1016/j.physrep.2018.11.002}}.

\bibitem{Buchmuller:1985jz}
W.~Buchmuller, D.~Wyler, {Effective Lagrangian Analysis of New Interactions and
  Flavor Conservation}, Nucl. Phys. B 268 (1986) 621--653.
\newblock \href {https://doi.org/10.1016/0550-3213(86)90262-2}
  {\path{doi:10.1016/0550-3213(86)90262-2}}.

\bibitem{Ellis:2018cos}
J.~Ellis, S.-F. Ge, {Constraining Gluonic Quartic Gauge Coupling Operators with
  gg\textrightarrow{}\ensuremath{\gamma}\ensuremath{\gamma}}, Phys. Rev. Lett.
  121~(4) (2018) 041801.
\newblock \href {http://arxiv.org/abs/1802.02416} {\path{arXiv:1802.02416}},
  \href {https://doi.org/10.1103/PhysRevLett.121.041801}
  {\path{doi:10.1103/PhysRevLett.121.041801}}.

\bibitem{Ellis:2019zex}
J.~Ellis, S.-F. Ge, H.-J. He, R.-Q. Xiao, {Probing the scale of new physics in
  the $ZZ\gamma$ coupling at $e^+e^-$ colliders}, Chin. Phys. C 44~(6) (2020)
  063106.
\newblock \href {http://arxiv.org/abs/1902.06631} {\path{arXiv:1902.06631}},
  \href {https://doi.org/10.1088/1674-1137/44/6/063106}
  {\path{doi:10.1088/1674-1137/44/6/063106}}.

\bibitem{Ellis:2020ljj}
J.~Ellis, H.-J. He, R.-Q. Xiao, {Probing new physics in dimension-8 neutral
  gauge couplings at e$^{+}$e$^{-}$ colliders}, Sci. China Phys. Mech. Astron.
  64~(2) (2021) 221062.
\newblock \href {http://arxiv.org/abs/2008.04298} {\path{arXiv:2008.04298}},
  \href {https://doi.org/10.1007/s11433-020-1617-3}
  {\path{doi:10.1007/s11433-020-1617-3}}.

\bibitem{Gounaris:2000dn}
G.~J. Gounaris, J.~Layssac, F.~M. Renard, {Off-shell structure of the anomalous
  $Z$ and $\gamma$ selfcouplings}, Phys. Rev. D 62 (2000) 073012.
\newblock \href {http://arxiv.org/abs/hep-ph/0005269}
  {\path{arXiv:hep-ph/0005269}}, \href
  {https://doi.org/10.1103/PhysRevD.65.017302}
  {\path{doi:10.1103/PhysRevD.65.017302}}.

\bibitem{Gounaris:1999kf}
G.~J. Gounaris, J.~Layssac, F.~M. Renard, {Signatures of the anomalous
  $Z_{\gamma}$ and $Z Z$ production at the lepton and hadron colliders}, Phys.
  Rev. D 61 (2000) 073013.
\newblock \href {http://arxiv.org/abs/hep-ph/9910395}
  {\path{arXiv:hep-ph/9910395}}, \href
  {https://doi.org/10.1103/PhysRevD.61.073013}
  {\path{doi:10.1103/PhysRevD.61.073013}}.

\bibitem{Senol:2018cks}
A.~Senol, H.~Denizli, A.~Yilmaz, I.~Turk~Cakir, K.~Y. Oyulmaz, O.~Karadeniz,
  O.~Cakir, {Probing the Effects of Dimension-eight Operators Describing
  Anomalous Neutral Triple Gauge Boson Interactions at FCC-hh}, Nucl. Phys. B
  935 (2018) 365--376.
\newblock \href {http://arxiv.org/abs/1805.03475} {\path{arXiv:1805.03475}},
  \href {https://doi.org/10.1016/j.nuclphysb.2018.08.018}
  {\path{doi:10.1016/j.nuclphysb.2018.08.018}}.

\bibitem{Fu:2021mub}
Q.~Fu, J.-C. Yang, C.-X. Yue, Y.-C. Guo, {The study of neutral triple gauge
  couplings in the process
  e+e\ensuremath{-}\textrightarrow{}Z\ensuremath{\gamma} including unitarity
  bounds}, Nucl. Phys. B 972 (2021) 115543.
\newblock \href {http://arxiv.org/abs/2102.03623} {\path{arXiv:2102.03623}},
  \href {https://doi.org/10.1016/j.nuclphysb.2021.115543}
  {\path{doi:10.1016/j.nuclphysb.2021.115543}}.

\bibitem{Degrande:2013kka}
C.~Degrande, {A basis of dimension-eight operators for anomalous neutral triple
  gauge boson interactions}, JHEP 02 (2014) 101.
\newblock \href {http://arxiv.org/abs/1308.6323} {\path{arXiv:1308.6323}},
  \href {https://doi.org/10.1007/JHEP02(2014)101}
  {\path{doi:10.1007/JHEP02(2014)101}}.

\bibitem{Jahedi:2022duc}
S.~Jahedi, J.~Lahiri, {Probing anomalous ZZ\ensuremath{\gamma} and
  Z\ensuremath{\gamma}\ensuremath{\gamma} couplings at the e$^{+}$e$^{-}$
  colliders using optimal observable technique}, JHEP 04 (2023) 085.
\newblock \href {http://arxiv.org/abs/2212.05121} {\path{arXiv:2212.05121}},
  \href {https://doi.org/10.1007/JHEP04(2023)085}
  {\path{doi:10.1007/JHEP04(2023)085}}.

\bibitem{Jahedi:2023myu}
S.~Jahedi, {Optimal estimation of dimension-8 neutral triple gauge couplings at
  the e$^{+}$e$^{-}$ colliders}, JHEP 12 (2023) 031.
\newblock \href {http://arxiv.org/abs/2305.11266} {\path{arXiv:2305.11266}},
  \href {https://doi.org/10.1007/JHEP12(2023)031}
  {\path{doi:10.1007/JHEP12(2023)031}}.

\bibitem{Ellis:2017edi}
J.~Ellis, N.~E. Mavromatos, T.~You, {Light-by-Light Scattering Constraint on
  Born-Infeld Theory}, Phys. Rev. Lett. 118~(26) (2017) 261802.
\newblock \href {http://arxiv.org/abs/1703.08450} {\path{arXiv:1703.08450}},
  \href {https://doi.org/10.1103/PhysRevLett.118.261802}
  {\path{doi:10.1103/PhysRevLett.118.261802}}.

\bibitem{Bi:2019phv}
Q.~Bi, C.~Zhang, S.-Y. Zhou, {Positivity constraints on aQGC: carving out the
  physical parameter space}, JHEP 06 (2019) 137.
\newblock \href {http://arxiv.org/abs/1902.08977} {\path{arXiv:1902.08977}},
  \href {https://doi.org/10.1007/JHEP06(2019)137}
  {\path{doi:10.1007/JHEP06(2019)137}}.

\bibitem{Zhang:2020jyn}
C.~Zhang, S.-Y. Zhou, {Convex Geometry Perspective on the (Standard Model)
  Effective Field Theory Space}, Phys. Rev. Lett. 125~(20) (2020) 201601.
\newblock \href {http://arxiv.org/abs/2005.03047} {\path{arXiv:2005.03047}},
  \href {https://doi.org/10.1103/PhysRevLett.125.201601}
  {\path{doi:10.1103/PhysRevLett.125.201601}}.

\bibitem{Yamashita:2020gtt}
K.~Yamashita, C.~Zhang, S.-Y. Zhou, {Elastic positivity vs extremal positivity
  bounds in SMEFT: a case study in transversal electroweak gauge-boson
  scatterings}, JHEP 01 (2021) 095.
\newblock \href {http://arxiv.org/abs/2009.04490} {\path{arXiv:2009.04490}},
  \href {https://doi.org/10.1007/JHEP01(2021)095}
  {\path{doi:10.1007/JHEP01(2021)095}}.

\bibitem{Zhang:2023khv}
Y.-T. Zhang, X.-T. Wang, J.-C. Yang, {Searching for gluon quartic gauge
  couplings at muon colliders using the autoencoder}, Phys. Rev. D 109~(9)
  (2024) 095028.
\newblock \href {http://arxiv.org/abs/2311.16627} {\path{arXiv:2311.16627}},
  \href {https://doi.org/10.1103/PhysRevD.109.095028}
  {\path{doi:10.1103/PhysRevD.109.095028}}.

\bibitem{Zhang:2023ykh}
S.~Zhang, Y.-C. Guo, J.-C. Yang, {Optimize the event selection strategy to
  study the anomalous quartic gauge couplings at muon colliders using the
  support vector machine and quantum support vector machine}, Eur. Phys. J. C
  84~(8) (2024) 833.
\newblock \href {http://arxiv.org/abs/2311.15280} {\path{arXiv:2311.15280}},
  \href {https://doi.org/10.1140/epjc/s10052-024-13208-4}
  {\path{doi:10.1140/epjc/s10052-024-13208-4}}.

\bibitem{Dong:2023nir}
Y.-F. Dong, Y.-C. Mao, i.-C. Yang, J.-C. Yang, {Searching for anomalous quartic
  gauge couplings at muon colliders using principal component analysis}, Eur.
  Phys. J. C 83~(7) (2023) 555.
\newblock \href {http://arxiv.org/abs/2304.01505} {\path{arXiv:2304.01505}},
  \href {https://doi.org/10.1140/epjc/s10052-023-11719-0}
  {\path{doi:10.1140/epjc/s10052-023-11719-0}}.

\bibitem{Zhang:2023yfg}
S.~Zhang, J.-C. Yang, Y.-C. Guo, {Using k-means assistant event selection
  strategy to study anomalous quartic gauge couplings at muon colliders}, Eur.
  Phys. J. C 84~(2) (2024) 142.
\newblock \href {http://arxiv.org/abs/2302.01274} {\path{arXiv:2302.01274}},
  \href {https://doi.org/10.1140/epjc/s10052-024-12494-2}
  {\path{doi:10.1140/epjc/s10052-024-12494-2}}.

\bibitem{Yang:2022fhw}
J.-C. Yang, X.-Y. Han, Z.-B. Qin, T.~Li, Y.-C. Guo, {Measuring the anomalous
  quartic gauge couplings in the W+W\ensuremath{-}\textrightarrow
  W+W\ensuremath{-} process at muon collider using artificial neural networks},
  JHEP 09 (2022) 074.
\newblock \href {http://arxiv.org/abs/2204.10034} {\path{arXiv:2204.10034}},
  \href {https://doi.org/10.1007/JHEP09(2022)074}
  {\path{doi:10.1007/JHEP09(2022)074}}.

\bibitem{Yang:2021kyy}
J.-C. Yang, Y.-C. Guo, L.-H. Cai, {Using a nested anomaly detection machine
  learning algorithm to study the neutral triple gauge couplings at an
  e+e\ensuremath{-} collider}, Nucl. Phys. B 977 (2022) 115735.
\newblock \href {http://arxiv.org/abs/2111.10543} {\path{arXiv:2111.10543}},
  \href {https://doi.org/10.1016/j.nuclphysb.2022.115735}
  {\path{doi:10.1016/j.nuclphysb.2022.115735}}.

\bibitem{Jiang:2021ytz}
L.~Jiang, Y.-C. Guo, J.-C. Yang, {Detecting anomalous quartic gauge couplings
  using the isolation forest machine learning algorithm}, Phys. Rev. D 104~(3)
  (2021) 035021.
\newblock \href {http://arxiv.org/abs/2103.03151} {\path{arXiv:2103.03151}},
  \href {https://doi.org/10.1103/PhysRevD.104.035021}
  {\path{doi:10.1103/PhysRevD.104.035021}}.

\bibitem{Vaslin:2023lig}
L.~Vaslin, V.~Barra, J.~Donini, {GAN-AE: an anomaly detection algorithm for New
  Physics search in LHC data}, Eur. Phys. J. C 83~(11) (2023) 1008.
\newblock \href {http://arxiv.org/abs/2305.15179} {\path{arXiv:2305.15179}},
  \href {https://doi.org/10.1140/epjc/s10052-023-12169-4}
  {\path{doi:10.1140/epjc/s10052-023-12169-4}}.

\bibitem{Kuusela:2011aa}
M.~Kuusela, T.~Vatanen, E.~Malmi, T.~Raiko, T.~Aaltonen, Y.~Nagai,
  {Semi-Supervised Anomaly Detection - Towards Model-Independent Searches of
  New Physics}, J. Phys. Conf. Ser. 368 (2012) 012032.
\newblock \href {http://arxiv.org/abs/1112.3329} {\path{arXiv:1112.3329}},
  \href {https://doi.org/10.1088/1742-6596/368/1/012032}
  {\path{doi:10.1088/1742-6596/368/1/012032}}.

\bibitem{Collins:2018epr}
J.~H. Collins, K.~Howe, B.~Nachman, {Anomaly Detection for Resonant New Physics
  with Machine Learning}, Phys. Rev. Lett. 121~(24) (2018) 241803.
\newblock \href {http://arxiv.org/abs/1805.02664} {\path{arXiv:1805.02664}},
  \href {https://doi.org/10.1103/PhysRevLett.121.241803}
  {\path{doi:10.1103/PhysRevLett.121.241803}}.

\bibitem{Atkinson:2022uzb}
O.~Atkinson, A.~Bhardwaj, C.~Englert, P.~Konar, V.~S. Ngairangbam,
  M.~Spannowsky, {IRC-Safe Graph Autoencoder for Unsupervised Anomaly
  Detection}, Front. Artif. Intell. 5 (2022) 943135.
\newblock \href {http://arxiv.org/abs/2204.12231} {\path{arXiv:2204.12231}},
  \href {https://doi.org/10.3389/frai.2022.943135}
  {\path{doi:10.3389/frai.2022.943135}}.

\bibitem{Kasieczka:2021xcg}
G.~Kasieczka, et~al., {The LHC Olympics 2020 a community challenge for anomaly
  detection in high energy physics}, Rept. Prog. Phys. 84~(12) (2021) 124201.
\newblock \href {http://arxiv.org/abs/2101.08320} {\path{arXiv:2101.08320}},
  \href {https://doi.org/10.1088/1361-6633/ac36b9}
  {\path{doi:10.1088/1361-6633/ac36b9}}.

\bibitem{Farina:2018fyg}
M.~Farina, Y.~Nakai, D.~Shih, {Searching for New Physics with Deep
  Autoencoders}, Phys. Rev. D 101~(7) (2020) 075021.
\newblock \href {http://arxiv.org/abs/1808.08992} {\path{arXiv:1808.08992}},
  \href {https://doi.org/10.1103/PhysRevD.101.075021}
  {\path{doi:10.1103/PhysRevD.101.075021}}.

\bibitem{Cerri:2018anq}
O.~Cerri, T.~Q. Nguyen, M.~Pierini, M.~Spiropulu, J.-R. Vlimant, {Variational
  Autoencoders for New Physics Mining at the Large Hadron Collider}, JHEP 05
  (2019) 036.
\newblock \href {http://arxiv.org/abs/1811.10276} {\path{arXiv:1811.10276}},
  \href {https://doi.org/10.1007/JHEP05(2019)036}
  {\path{doi:10.1007/JHEP05(2019)036}}.

\bibitem{vanBeekveld:2020txa}
M.~van Beekveld, S.~Caron, L.~Hendriks, P.~Jackson, A.~Leinweber, S.~Otten,
  R.~Patrick, R.~Ruiz De~Austri, M.~Santoni, M.~White, {Combining outlier
  analysis algorithms to identify new physics at the LHC}, JHEP 09 (2021) 024.
\newblock \href {http://arxiv.org/abs/2010.07940} {\path{arXiv:2010.07940}},
  \href {https://doi.org/10.1007/JHEP09(2021)024}
  {\path{doi:10.1007/JHEP09(2021)024}}.

\bibitem{CrispimRomao:2020ucc}
M.~Crispim Rom\~ao, N.~F. Castro, R.~Pedro, {Finding New Physics without
  learning about it: Anomaly Detection as a tool for Searches at Colliders},
  Eur. Phys. J. C 81~(1) (2021) 27, [Erratum: Eur.Phys.J.C 81, 1020 (2021)].
\newblock \href {http://arxiv.org/abs/2006.05432} {\path{arXiv:2006.05432}},
  \href {https://doi.org/10.1140/epjc/s10052-021-09813-2}
  {\path{doi:10.1140/epjc/s10052-021-09813-2}}.

\bibitem{Ren:2017ymm}
J.~Ren, L.~Wu, J.~M. Yang, J.~Zhao, {Exploring supersymmetry with machine
  learning}, Nucl. Phys. B 943 (2019) 114613.
\newblock \href {http://arxiv.org/abs/1708.06615} {\path{arXiv:1708.06615}},
  \href {https://doi.org/10.1016/j.nuclphysb.2019.114613}
  {\path{doi:10.1016/j.nuclphysb.2019.114613}}.

\bibitem{Abdughani:2018wrw}
M.~Abdughani, J.~Ren, L.~Wu, J.~M. Yang, {Probing stop pair production at the
  LHC with graph neural networks}, JHEP 2019~(8) (2019) 055.
\newblock \href {http://arxiv.org/abs/1807.09088} {\path{arXiv:1807.09088}},
  \href {https://doi.org/10.1007/JHEP08(2019)055}
  {\path{doi:10.1007/JHEP08(2019)055}}.

\bibitem{Ren:2019xhp}
J.~Ren, L.~Wu, J.~M. Yang, {Unveiling CP property of top-Higgs coupling with
  graph neural networks at the LHC}, Phys. Lett. B 802 (2020) 135198.
\newblock \href {http://arxiv.org/abs/1901.05627} {\path{arXiv:1901.05627}},
  \href {https://doi.org/10.1016/j.physletb.2020.135198}
  {\path{doi:10.1016/j.physletb.2020.135198}}.

\bibitem{Liu:2022jsp}
W.~Liu, H.-W. Yin, Z.-R. Wang, W.-Q. Fan, {Multi-state Swap Test Algorithm},
  2022.
\newblock \href {http://arxiv.org/abs/2205.07171} {\path{arXiv:2205.07171}}.

\bibitem{Fanizza:2020qjq}
M.~Fanizza, M.~Rosati, M.~Skotiniotis, J.~Calsamiglia, V.~Giovannetti, {Beyond
  the Swap Test: Optimal Estimation of Quantum State Overlap}, Phys. Rev. Lett.
  124~(6) (2020) 060503.
\newblock \href {https://doi.org/10.1103/PhysRevLett.124.060503}
  {\path{doi:10.1103/PhysRevLett.124.060503}}.

\bibitem{Liu:2020lhd}
Y.~Liu, S.~Arunachalam, K.~Temme, {A rigorous and robust quantum speed-up in
  supervised machine learning}, Nature Phys. 17~(9) (2021) 1013--1017.
\newblock \href {http://arxiv.org/abs/2010.02174} {\path{arXiv:2010.02174}},
  \href {https://doi.org/10.1038/s41567-021-01287-z}
  {\path{doi:10.1038/s41567-021-01287-z}}.

\bibitem{Green:2016trm}
D.~R. Green, P.~Meade, M.-A. Pleier, {Multiboson interactions at the LHC}, Rev.
  Mod. Phys. 89~(3) (2017) 035008.
\newblock \href {http://arxiv.org/abs/1610.07572} {\path{arXiv:1610.07572}},
  \href {https://doi.org/10.1103/RevModPhys.89.035008}
  {\path{doi:10.1103/RevModPhys.89.035008}}.

\bibitem{Chang:2013aya}
J.~Chang, K.~Cheung, C.-T. Lu, T.-C. Yuan, {WW scattering in the era of
  post-Higgs-boson discovery}, Phys. Rev. D 87 (2013) 093005.
\newblock \href {http://arxiv.org/abs/1303.6335} {\path{arXiv:1303.6335}},
  \href {https://doi.org/10.1103/PhysRevD.87.093005}
  {\path{doi:10.1103/PhysRevD.87.093005}}.

\bibitem{Anders:2018oin}
C.~F. Anders, et~al., {Vector boson scattering: Recent experimental and theory
  developments}, Rev. Phys. 3 (2018) 44--63.
\newblock \href {http://arxiv.org/abs/1801.04203} {\path{arXiv:1801.04203}},
  \href {https://doi.org/10.1016/j.revip.2018.11.001}
  {\path{doi:10.1016/j.revip.2018.11.001}}.

\bibitem{Zhang:2018shp}
C.~Zhang, S.-Y. Zhou, {Positivity bounds on vector boson scattering at the
  LHC}, Phys. Rev. D 100~(9) (2019) 095003.
\newblock \href {http://arxiv.org/abs/1808.00010} {\path{arXiv:1808.00010}},
  \href {https://doi.org/10.1103/PhysRevD.100.095003}
  {\path{doi:10.1103/PhysRevD.100.095003}}.

\bibitem{Guo:2020lim}
Y.-C. Guo, Y.-Y. Wang, J.-C. Yang, C.-X. Yue, {Constraints on anomalous quartic
  gauge couplings via $W\gamma jj$ production at the LHC}, Chin. Phys. C
  44~(12) (2020) 123105.
\newblock \href {http://arxiv.org/abs/2002.03326} {\path{arXiv:2002.03326}},
  \href {https://doi.org/10.1088/1674-1137/abb4d2}
  {\path{doi:10.1088/1674-1137/abb4d2}}.

\bibitem{Guo:2019agy}
Y.-C. Guo, Y.-Y. Wang, J.-C. Yang, {Constraints on anomalous quartic gauge
  couplings by $\gamma\gamma \to W^+W^-$ scattering}, Nucl. Phys. B 961 (2020)
  115222.
\newblock \href {http://arxiv.org/abs/1912.10686} {\path{arXiv:1912.10686}},
  \href {https://doi.org/10.1016/j.nuclphysb.2020.115222}
  {\path{doi:10.1016/j.nuclphysb.2020.115222}}.

\bibitem{Yang:2021pcf}
J.-C. Yang, Y.-C. Guo, C.-X. Yue, Q.~Fu, {Constraints on anomalous quartic
  gauge couplings via Z\ensuremath{\gamma}jj production at the LHC}, Phys. Rev.
  D 104~(3) (2021) 035015.
\newblock \href {http://arxiv.org/abs/2107.01123} {\path{arXiv:2107.01123}},
  \href {https://doi.org/10.1103/PhysRevD.104.035015}
  {\path{doi:10.1103/PhysRevD.104.035015}}.

\bibitem{Yang:2020rjt}
J.-C. Yang, Z.-B. Qing, X.-Y. Han, Y.-C. Guo, T.~Li, {Tri-photon at muon
  collider: a new process to probe the anomalous quartic gauge couplings}, JHEP
  22 (2020) 053.
\newblock \href {http://arxiv.org/abs/2204.08195} {\path{arXiv:2204.08195}},
  \href {https://doi.org/10.1007/JHEP07(2022)053}
  {\path{doi:10.1007/JHEP07(2022)053}}.

\bibitem{ATLAS:2014jzl}
G.~Aad, et~al., {Evidence for Electroweak Production of $W^{\pm}W^{\pm}jj$ in
  $pp$ Collisions at $\sqrt{s}=8$ TeV with the ATLAS Detector}, Phys. Rev.
  Lett. 113~(14) (2014) 141803.
\newblock \href {http://arxiv.org/abs/1405.6241} {\path{arXiv:1405.6241}},
  \href {https://doi.org/10.1103/PhysRevLett.113.141803}
  {\path{doi:10.1103/PhysRevLett.113.141803}}.

\bibitem{CMS:2020gfh}
A.~M. Sirunyan, et~al., {Measurements of production cross sections of WZ and
  same-sign WW boson pairs in association with two jets in proton-proton
  collisions at $\sqrt{s} =$ 13 TeV}, Phys. Lett. B 809 (2020) 135710.
\newblock \href {http://arxiv.org/abs/2005.01173} {\path{arXiv:2005.01173}},
  \href {https://doi.org/10.1016/j.physletb.2020.135710}
  {\path{doi:10.1016/j.physletb.2020.135710}}.

\bibitem{ATLAS:2017vqm}
M.~Aaboud, et~al., {Studies of $Z\gamma$ production in association with a
  high-mass dijet system in $pp$ collisions at $\sqrt{s}=$ 8 TeV with the ATLAS
  detector}, JHEP 2017~(7) (2017) 107.
\newblock \href {http://arxiv.org/abs/1705.01966} {\path{arXiv:1705.01966}},
  \href {https://doi.org/10.1007/JHEP07(2017)107}
  {\path{doi:10.1007/JHEP07(2017)107}}.

\bibitem{CMS:2017rin}
V.~Khachatryan, et~al., {Measurement of the cross section for electroweak
  production of Z$\gamma$ in association with two jets and constraints on
  anomalous quartic gauge couplings in proton\textendash{}proton collisions at
  $\sqrt{s} = 8$ TeV}, Phys. Lett. B 770 (2017) 380--402.
\newblock \href {http://arxiv.org/abs/1702.03025} {\path{arXiv:1702.03025}},
  \href {https://doi.org/10.1016/j.physletb.2017.04.071}
  {\path{doi:10.1016/j.physletb.2017.04.071}}.

\bibitem{CMS:2020ioi}
A.~M. Sirunyan, et~al., {Measurement of the cross section for electroweak
  production of a Z boson, a photon and two jets in proton-proton collisions at
  $\sqrt{s} =$ 13 TeV and constraints on anomalous quartic couplings}, JHEP
  2020~(6) (2020) 76.
\newblock \href {http://arxiv.org/abs/2002.09902} {\path{arXiv:2002.09902}},
  \href {https://doi.org/10.1007/JHEP06(2020)076}
  {\path{doi:10.1007/JHEP06(2020)076}}.

\bibitem{CMS:2016gct}
V.~Khachatryan, et~al., {Measurement of electroweak-induced production of
  W$\gamma$ with two jets in pp collisions at $ \sqrt{s}=8 $ TeV and
  constraints on anomalous quartic gauge couplings}, JHEP 2017~(6) (2017) 106.
\newblock \href {http://arxiv.org/abs/1612.09256} {\path{arXiv:1612.09256}},
  \href {https://doi.org/10.1007/JHEP06(2017)106}
  {\path{doi:10.1007/JHEP06(2017)106}}.

\bibitem{CMS:2017zmo}
A.~M. Sirunyan, et~al., {Measurement of vector boson scattering and constraints
  on anomalous quartic couplings from events with four leptons and two jets in
  proton\textendash{}proton collisions at $\sqrt{s}=$ 13 TeV}, Phys. Lett. B
  774 (2017) 682--705.
\newblock \href {http://arxiv.org/abs/1708.02812} {\path{arXiv:1708.02812}},
  \href {https://doi.org/10.1016/j.physletb.2017.10.020}
  {\path{doi:10.1016/j.physletb.2017.10.020}}.

\bibitem{CMS:2018ccg}
A.~M. Sirunyan, et~al., {Measurement of differential cross sections for Z boson
  pair production in association with jets at $\sqrt{s} =$ 8 and 13 TeV}, Phys.
  Lett. B 789 (2019) 19--44.
\newblock \href {http://arxiv.org/abs/1806.11073} {\path{arXiv:1806.11073}},
  \href {https://doi.org/10.1016/j.physletb.2018.11.007}
  {\path{doi:10.1016/j.physletb.2018.11.007}}.

\bibitem{ATLAS:2018mxa}
M.~Aaboud, et~al., {Observation of electroweak $W^{\pm}Z$ boson pair production
  in association with two jets in $pp$ collisions at $\sqrt{s} =$ 13 TeV with
  the ATLAS detector}, Phys. Lett. B 793 (2019) 469--492.
\newblock \href {http://arxiv.org/abs/1812.09740} {\path{arXiv:1812.09740}},
  \href {https://doi.org/10.1016/j.physletb.2019.05.012}
  {\path{doi:10.1016/j.physletb.2019.05.012}}.

\bibitem{CMS:2019uys}
A.~M. Sirunyan, et~al., {Measurement of electroweak WZ boson production and
  search for new physics in WZ + two jets events in pp collisions at $\sqrt{s}
  =$ 13TeV}, Phys. Lett. B 795 (2019) 281--307.
\newblock \href {http://arxiv.org/abs/1901.04060} {\path{arXiv:1901.04060}},
  \href {https://doi.org/10.1016/j.physletb.2019.05.042}
  {\path{doi:10.1016/j.physletb.2019.05.042}}.

\bibitem{CMS:2016rtz}
V.~Khachatryan, et~al., {Evidence for exclusive $\gamma\gamma \to W^+ W^-$
  production and constraints on anomalous quartic gauge couplings in $pp$
  collisions at $ \sqrt{s}=7 $ and 8 TeV}, JHEP 2016~(8) (2016) 119.
\newblock \href {http://arxiv.org/abs/1604.04464} {\path{arXiv:1604.04464}},
  \href {https://doi.org/10.1007/JHEP08(2016)119}
  {\path{doi:10.1007/JHEP08(2016)119}}.

\bibitem{CMS:2017fhs}
A.~M. Sirunyan, et~al., {Observation of electroweak production of same-sign W
  boson pairs in the two jet and two same-sign lepton final state in
  proton-proton collisions at $\sqrt{s} = $ 13 TeV}, Phys. Rev. Lett. 120~(8)
  (2018) 081801.
\newblock \href {http://arxiv.org/abs/1709.05822} {\path{arXiv:1709.05822}},
  \href {https://doi.org/10.1103/PhysRevLett.120.081801}
  {\path{doi:10.1103/PhysRevLett.120.081801}}.

\bibitem{CMS:2019qfk}
A.~M. Sirunyan, et~al., {Search for anomalous electroweak production of vector
  boson pairs in association with two jets in proton-proton collisions at 13
  TeV}, Phys. Lett. B 798 (2019) 134985.
\newblock \href {http://arxiv.org/abs/1905.07445} {\path{arXiv:1905.07445}},
  \href {https://doi.org/10.1016/j.physletb.2019.134985}
  {\path{doi:10.1016/j.physletb.2019.134985}}.

\bibitem{CMS:2020ypo}
A.~M. Sirunyan, et~al., {Observation of electroweak production of W$\gamma$
  with two jets in proton-proton collisions at $\sqrt {s}$ = 13 TeV}, Phys.
  Lett. B 811 (2020) 135988.
\newblock \href {http://arxiv.org/abs/2008.10521} {\path{arXiv:2008.10521}},
  \href {https://doi.org/10.1016/j.physletb.2020.135988}
  {\path{doi:10.1016/j.physletb.2020.135988}}.

\bibitem{CMS:2020fqz}
A.~M. Sirunyan, et~al., {Evidence for electroweak production of four charged
  leptons and two jets in proton-proton collisions at $\sqrt {s}$ = 13 TeV},
  Phys. Lett. B 812 (2021) 135992.
\newblock \href {http://arxiv.org/abs/2008.07013} {\path{arXiv:2008.07013}},
  \href {https://doi.org/10.1016/j.physletb.2020.135992}
  {\path{doi:10.1016/j.physletb.2020.135992}}.

\bibitem{Buttazzo:2018qqp}
D.~Buttazzo, D.~Redigolo, F.~Sala, A.~Tesi, {Fusing Vectors into Scalars at
  High Energy Lepton Colliders}, JHEP 11 (2018) 144.
\newblock \href {http://arxiv.org/abs/1807.04743} {\path{arXiv:1807.04743}},
  \href {https://doi.org/10.1007/JHEP11(2018)144}
  {\path{doi:10.1007/JHEP11(2018)144}}.

\bibitem{Delahaye:2019omf}
J.~P. Delahaye, M.~Diemoz, K.~Long, B.~Mansouli\'e, N.~Pastrone, L.~Rivkin,
  D.~Schulte, A.~Skrinsky, A.~Wulzer, {Muon Colliders}, 2019.
\newblock \href {http://arxiv.org/abs/1901.06150} {\path{arXiv:1901.06150}}.

\bibitem{Costantini:2020stv}
A.~Costantini, F.~De~Lillo, F.~Maltoni, L.~Mantani, O.~Mattelaer, R.~Ruiz,
  X.~Zhao, {Vector boson fusion at multi-TeV muon colliders}, JHEP 2020~(9)
  (2020) 080.
\newblock \href {http://arxiv.org/abs/2005.10289} {\path{arXiv:2005.10289}},
  \href {https://doi.org/10.1007/JHEP09(2020)080}
  {\path{doi:10.1007/JHEP09(2020)080}}.

\bibitem{Lu:2020dkx}
M.~Lu, A.~M. Levin, C.~Li, A.~Agapitos, Q.~Li, F.~Meng, S.~Qian, J.~Xiao,
  T.~Yang, {The physics case for an electron-muon collider}, Adv. High Energy
  Phys. 2021 (2021) 6693618.
\newblock \href {http://arxiv.org/abs/2010.15144} {\path{arXiv:2010.15144}},
  \href {https://doi.org/10.1155/2021/6693618}
  {\path{doi:10.1155/2021/6693618}}.

\bibitem{AlAli:2021let}
H.~Al~Ali, et~al., {The muon Smasher\textquoteright{}s guide}, Rept. Prog.
  Phys. 85~(8) (2022) 084201.
\newblock \href {http://arxiv.org/abs/2103.14043} {\path{arXiv:2103.14043}},
  \href {https://doi.org/10.1088/1361-6633/ac6678}
  {\path{doi:10.1088/1361-6633/ac6678}}.

\bibitem{Franceschini:2021aqd}
R.~Franceschini, M.~Greco, {Higgs and BSM Physics at the Future Muon Collider},
  Symmetry 13~(5) (2021) 851.
\newblock \href {http://arxiv.org/abs/2104.05770} {\path{arXiv:2104.05770}},
  \href {https://doi.org/10.3390/sym13050851} {\path{doi:10.3390/sym13050851}}.

\bibitem{Palmer:1996gs}
R.~Palmer, et~al., {Muon collider design}, Nucl. Phys. B Proc. Suppl. 51 (1996)
  61--84.
\newblock \href {http://arxiv.org/abs/acc-phys/9604001}
  {\path{arXiv:acc-phys/9604001}}, \href
  {https://doi.org/10.1016/0920-5632(96)00417-3}
  {\path{doi:10.1016/0920-5632(96)00417-3}}.

\bibitem{Holmes:2012aei}
S.~D. Holmes, V.~D. Shiltsev, {Muon Collider}, Springer-Verlag Berlin
  Heidelberg, Germany, 2013, pp. 816--822.
\newblock \href {http://arxiv.org/abs/1202.3803} {\path{arXiv:1202.3803}},
  \href {https://doi.org/10.1007/978-3-642-23053-0_48}
  {\path{doi:10.1007/978-3-642-23053-0_48}}.

\bibitem{Liu:2021jyc}
W.~Liu, K.-P. Xie, {Probing electroweak phase transition with multi-TeV muon
  colliders and gravitational waves}, JHEP 2021~(4) (2021) 015.
\newblock \href {http://arxiv.org/abs/2101.10469} {\path{arXiv:2101.10469}},
  \href {https://doi.org/10.1007/JHEP04(2021)015}
  {\path{doi:10.1007/JHEP04(2021)015}}.

\bibitem{Liu:2021akf}
W.~Liu, K.-P. Xie, Z.~Yi, {Testing leptogenesis at the LHC and future muon
  colliders: A Z' scenario}, Phys. Rev. D 105~(9) (2022) 095034.
\newblock \href {http://arxiv.org/abs/2109.15087} {\path{arXiv:2109.15087}},
  \href {https://doi.org/10.1103/PhysRevD.105.095034}
  {\path{doi:10.1103/PhysRevD.105.095034}}.

\bibitem{Eboli:2006wa}
O.~J.~P. Eboli, M.~C. Gonzalez-Garcia, J.~K. Mizukoshi, {p p ---\ensuremath{>}
  j j e+- mu+- nu nu and j j e+- mu-+ nu nu at O( alpha(em)**6) and
  O(alpha(em)**4 alpha(s)**2) for the study of the quartic electroweak gauge
  boson vertex at CERN LHC}, Phys. Rev. D 74 (2006) 073005.
\newblock \href {http://arxiv.org/abs/hep-ph/0606118}
  {\path{arXiv:hep-ph/0606118}}, \href
  {https://doi.org/10.1103/PhysRevD.74.073005}
  {\path{doi:10.1103/PhysRevD.74.073005}}.

\bibitem{Eboli:2016kko}
O.~J.~P. \'Eboli, M.~C. Gonzalez-Garcia, {Classifying the bosonic quartic
  couplings}, Phys. Rev. D 93~(9) (2016) 093013.
\newblock \href {http://arxiv.org/abs/1604.03555} {\path{arXiv:1604.03555}},
  \href {https://doi.org/10.1103/PhysRevD.93.093013}
  {\path{doi:10.1103/PhysRevD.93.093013}}.

\bibitem{Yue:2021snv}
C.-X. Yue, X.-J. Cheng, J.-C. Yang, {Charged-current non-standard neutrino
  interactions at the LHC and HL-LHC*}, Chin. Phys. C 47~(4) (2023) 043111.
\newblock \href {http://arxiv.org/abs/2110.01204} {\path{arXiv:2110.01204}},
  \href {https://doi.org/10.1088/1674-1137/acb993}
  {\path{doi:10.1088/1674-1137/acb993}}.

\bibitem{Yang:2022ilt}
J.-C. Yang, Y.-C. Guo, B.~Liu, T.~Li, {Shining light on magnetic monopoles
  through high-energy muon colliders}, Nucl. Phys. B 987 (2023) 116097.
\newblock \href {http://arxiv.org/abs/2208.02188} {\path{arXiv:2208.02188}},
  \href {https://doi.org/10.1016/j.nuclphysb.2023.116097}
  {\path{doi:10.1016/j.nuclphysb.2023.116097}}.

\bibitem{Layssac:1993vfp}
J.~Layssac, F.~M. Renard, G.~J. Gounaris, {Unitarity constraints for transverse
  gauge bosons at LEP and supercolliders}, Phys. Lett. B 332 (1994) 146--152.
\newblock \href {http://arxiv.org/abs/hep-ph/9311370}
  {\path{arXiv:hep-ph/9311370}}, \href
  {https://doi.org/10.1016/0370-2693(94)90872-9}
  {\path{doi:10.1016/0370-2693(94)90872-9}}.

\bibitem{Corbett:2017qgl}
T.~Corbett, O.~J.~P. \'Eboli, M.~C. Gonzalez-Garcia, {Unitarity Constraints on
  Dimension-six Operators II: Including Fermionic Operators}, Phys. Rev. D
  96~(3) (2017) 035006.
\newblock \href {http://arxiv.org/abs/1705.09294} {\path{arXiv:1705.09294}},
  \href {https://doi.org/10.1103/PhysRevD.96.035006}
  {\path{doi:10.1103/PhysRevD.96.035006}}.

\bibitem{Almeida:2020ylr}
E.~d.~S. Almeida, O.~J.~P. \'Eboli, M.~C. Gonzalez\textendash{}Garcia,
  {Unitarity constraints on anomalous quartic couplings}, Phys. Rev. D 101~(11)
  (2020) 113003.
\newblock \href {http://arxiv.org/abs/2004.05174} {\path{arXiv:2004.05174}},
  \href {https://doi.org/10.1103/PhysRevD.101.113003}
  {\path{doi:10.1103/PhysRevD.101.113003}}.

\bibitem{Kilian:2018bhs}
W.~Kilian, S.~Sun, Q.-S. Yan, X.~Zhao, Z.~Zhao, {Multi-Higgs boson production
  and unitarity in vector-boson fusion at future hadron colliders}, Phys. Rev.
  D 101~(7) (2020) 076012.
\newblock \href {http://arxiv.org/abs/1808.05534} {\path{arXiv:1808.05534}},
  \href {https://doi.org/10.1103/PhysRevD.101.076012}
  {\path{doi:10.1103/PhysRevD.101.076012}}.

\bibitem{Kilian:2021whd}
W.~Kilian, S.~Sun, Q.-S. Yan, X.~Zhao, Z.~Zhao, {Highly Boosted Higgs Bosons
  and Unitarity in Vector-Boson Fusion at Future Hadron Colliders}, JHEP 05
  (2021) 198.
\newblock \href {http://arxiv.org/abs/2101.12537} {\path{arXiv:2101.12537}},
  \href {https://doi.org/10.1007/JHEP05(2021)198}
  {\path{doi:10.1007/JHEP05(2021)198}}.

\bibitem{Perez:2018kav}
G.~Perez, M.~Sekulla, D.~Zeppenfeld, {Anomalous quartic gauge couplings and
  unitarization for the vector boson scattering process $pp\rightarrow
  W^+W^+jjX\rightarrow \ell ^+\nu _\ell \ell ^+\nu _\ell jjX$}, Eur. Phys. J. C
  78~(9) (2018) 759.
\newblock \href {http://arxiv.org/abs/1807.02707} {\path{arXiv:1807.02707}},
  \href {https://doi.org/10.1140/epjc/s10052-018-6230-1}
  {\path{doi:10.1140/epjc/s10052-018-6230-1}}.

\bibitem{Jacob:1959at}
M.~Jacob, G.~C. Wick, {On the General Theory of Collisions for Particles with
  Spin}, Annals Phys. 7 (1959) 404--428.
\newblock \href {https://doi.org/10.1006/aphy.2000.6022}
  {\path{doi:10.1006/aphy.2000.6022}}.

\bibitem{Corbett:2014ora}
T.~Corbett, O.~J.~P. \'Eboli, M.~C. Gonzalez-Garcia, {Unitarity Constraints on
  Dimension-Six Operators}, Phys. Rev. D 91~(3) (2015) 035014.
\newblock \href {http://arxiv.org/abs/1411.5026} {\path{arXiv:1411.5026}},
  \href {https://doi.org/10.1103/PhysRevD.91.035014}
  {\path{doi:10.1103/PhysRevD.91.035014}}.

\bibitem{Yang:2021ukg}
J.-C. Yang, J.-H. Chen, Y.-C. Guo, {Extract the energy scale of anomalous
  \ensuremath{\gamma}\ensuremath{\gamma} \textrightarrow{} W+W\ensuremath{-}
  scattering in the vector boson scattering process using artificial neural
  networks}, JHEP 09 (2021) 085.
\newblock \href {http://arxiv.org/abs/2107.13624} {\path{arXiv:2107.13624}},
  \href {https://doi.org/10.1007/JHEP09(2021)085}
  {\path{doi:10.1007/JHEP09(2021)085}}.

\bibitem{Chen:2022yiu}
M.~Chen, D.~Liu, {Top Yukawa coupling measurement at the muon collider}, Phys.
  Rev. D 109~(7) (2024) 075020.
\newblock \href {http://arxiv.org/abs/2212.11067} {\path{arXiv:2212.11067}},
  \href {https://doi.org/10.1103/PhysRevD.109.075020}
  {\path{doi:10.1103/PhysRevD.109.075020}}.

\bibitem{Han:2024gan}
T.~Han, D.~Liu, S.~Wang, {Top quark electroweak dipole moment at a high energy
  muon collider}, Phys. Rev. D 111~(3) (2025) 035015.
\newblock \href {http://arxiv.org/abs/2410.11015} {\path{arXiv:2410.11015}},
  \href {https://doi.org/10.1103/PhysRevD.111.035015}
  {\path{doi:10.1103/PhysRevD.111.035015}}.

\bibitem{Apollinari:2015wtw}
G.~Apollinari, O.~Br\"uning, T.~Nakamoto, L.~Rossi, {High Luminosity Large
  Hadron Collider HL-LHC}, CERN Yellow Rep.~(5) (2015) 1--19.
\newblock \href {http://arxiv.org/abs/1705.08830} {\path{arXiv:1705.08830}},
  \href {https://doi.org/10.5170/CERN-2015-005.1}
  {\path{doi:10.5170/CERN-2015-005.1}}.

\bibitem{Guan:2020bdl}
W.~Guan, G.~Perdue, A.~Pesah, M.~Schuld, K.~Terashi, S.~Vallecorsa, J.-R.
  Vlimant, {Quantum Machine Learning in High Energy Physics}, Mach. Learn. Sci.
  Tech. 2 (2021) 011003.
\newblock \href {http://arxiv.org/abs/2005.08582} {\path{arXiv:2005.08582}},
  \href {https://doi.org/10.1088/2632-2153/abc17d}
  {\path{doi:10.1088/2632-2153/abc17d}}.

\bibitem{Wu:2021xsj}
S.~L. Wu, et~al., {Application of quantum machine learning using the quantum
  kernel algorithm on high energy physics analysis at the LHC}, Phys. Rev. Res.
  3~(3) (2021) 033221.
\newblock \href {http://arxiv.org/abs/2104.05059} {\path{arXiv:2104.05059}},
  \href {https://doi.org/10.1103/PhysRevResearch.3.033221}
  {\path{doi:10.1103/PhysRevResearch.3.033221}}.

\bibitem{Wu:2020cye}
S.~L. Wu, et~al., {Application of quantum machine learning using the quantum
  variational classifier method to high energy physics analysis at the LHC on
  IBM quantum computer simulator and hardware with 10 qubits}, J. Phys. G
  48~(12) (2021) 125003.
\newblock \href {http://arxiv.org/abs/2012.11560} {\path{arXiv:2012.11560}},
  \href {https://doi.org/10.1088/1361-6471/ac1391}
  {\path{doi:10.1088/1361-6471/ac1391}}.

\bibitem{Terashi:2020wfi}
K.~Terashi, M.~Kaneda, T.~Kishimoto, M.~Saito, R.~Sawada, J.~Tanaka, {Event
  Classification with Quantum Machine Learning in High-Energy Physics}, Comput.
  Softw. Big Sci. 5~(1) (2021) 2.
\newblock \href {http://arxiv.org/abs/2002.09935} {\path{arXiv:2002.09935}},
  \href {https://doi.org/10.1007/s41781-020-00047-7}
  {\path{doi:10.1007/s41781-020-00047-7}}.

\bibitem{Havlicek:2018nqz}
V.~Havlicek, A.~D. C\'orcoles, K.~Temme, A.~W. Harrow, A.~Kandala, J.~M. Chow,
  J.~M. Gambetta, {Supervised learning with quantum-enhanced feature spaces},
  Nature 567 (2019) 209--212.
\newblock \href {http://arxiv.org/abs/1804.11326} {\path{arXiv:1804.11326}},
  \href {https://doi.org/10.1038/s41586-019-0980-2}
  {\path{doi:10.1038/s41586-019-0980-2}}.

\bibitem{Sherstov:2020qax}
A.~A. Sherstov, A.~A. Storozhenko, P.~Wu, {An optimal separation of randomized
  and Quantum query complexity}, SIAM J. Comput. 52~(2) (2023) 525--567.
\newblock \href {http://arxiv.org/abs/2008.10223} {\path{arXiv:2008.10223}},
  \href {https://doi.org/10.1145/3406325.3451019}
  {\path{doi:10.1145/3406325.3451019}}.

\bibitem{Alloul:2013bka}
A.~Alloul, N.~D. Christensen, C.~Degrande, C.~Duhr, B.~Fuks, {FeynRules 2.0 - A
  complete toolbox for tree-level phenomenology}, Comput. Phys. Commun. 185
  (2014) 2250--2300.
\newblock \href {http://arxiv.org/abs/1310.1921} {\path{arXiv:1310.1921}},
  \href {https://doi.org/10.1016/j.cpc.2014.04.012}
  {\path{doi:10.1016/j.cpc.2014.04.012}}.

\bibitem{Alwall:2014hca}
J.~Alwall, R.~Frederix, S.~Frixione, V.~Hirschi, F.~Maltoni, O.~Mattelaer,
  H.~S. Shao, T.~Stelzer, P.~Torrielli, M.~Zaro, {The automated computation of
  tree-level and next-to-leading order differential cross sections, and their
  matching to parton shower simulations}, JHEP 2014~(7) (2014) 079.
\newblock \href {http://arxiv.org/abs/1405.0301} {\path{arXiv:1405.0301}},
  \href {https://doi.org/10.1007/JHEP07(2014)079}
  {\path{doi:10.1007/JHEP07(2014)079}}.

\bibitem{Christensen:2008py}
N.~D. Christensen, C.~Duhr, {FeynRules - Feynman rules made easy}, Comput.
  Phys. Commun. 180 (2009) 1614--1641.
\newblock \href {http://arxiv.org/abs/0806.4194} {\path{arXiv:0806.4194}},
  \href {https://doi.org/10.1016/j.cpc.2009.02.018}
  {\path{doi:10.1016/j.cpc.2009.02.018}}.

\bibitem{Degrande:2011ua}
C.~Degrande, C.~Duhr, B.~Fuks, D.~Grellscheid, O.~Mattelaer, T.~Reiter, {UFO -
  The Universal FeynRules Output}, Comput. Phys. Commun. 183 (2012) 1201--1214.
\newblock \href {http://arxiv.org/abs/1108.2040} {\path{arXiv:1108.2040}},
  \href {https://doi.org/10.1016/j.cpc.2012.01.022}
  {\path{doi:10.1016/j.cpc.2012.01.022}}.

\bibitem{deFavereau:2013fsa}
J.~de~Favereau, C.~Delaere, P.~Demin, A.~Giammanco, V.~Lema\^\i{}tre,
  A.~Mertens, M.~Selvaggi, {DELPHES 3, A modular framework for fast simulation
  of a generic collider experiment}, JHEP 02 (2014) 057.
\newblock \href {http://arxiv.org/abs/1307.6346} {\path{arXiv:1307.6346}},
  \href {https://doi.org/10.1007/JHEP02(2014)057}
  {\path{doi:10.1007/JHEP02(2014)057}}.

\bibitem{MLAnalysis}
Y.-C. Guo, F.~Feng, A.~Di, S.-Q. Lu, J.-C. Yang, {MLAnalysis: An open-source
  program for high energy physics analyses}, Comput. Phys. Commun. 294 (2024)
  108957.
\newblock \href {http://arxiv.org/abs/2305.00964} {\path{arXiv:2305.00964}},
  \href {https://doi.org/10.1016/j.cpc.2023.108957}
  {\path{doi:10.1016/j.cpc.2023.108957}}.

\bibitem{Donoho_2004}
D.~Donoho, J.~Jin, \href{http://dx.doi.org/10.1214/009053604000000265}{Higher
  criticism for detecting sparse heterogeneous mixtures}, The Annals of
  Statistics 32~(3) (Jun. 2004).
\newblock \href {https://doi.org/10.1214/009053604000000265}
  {\path{doi:10.1214/009053604000000265}}.
\newline\urlprefix\url{http://dx.doi.org/10.1214/009053604000000265}

\bibitem{JMLR:v21:20-091}
R.~Tavenard, J.~Faouzi, G.~Vandewiele, F.~Divo, G.~Androz, C.~Holtz, M.~Payne,
  R.~Yurchak, M.~Ru{\ss}wurm, K.~Kolar, E.~Woods,
  \href{http://jmlr.org/papers/v21/20-091.html}{Tslearn, a machine learning
  toolkit for time series data}, Journal of Machine Learning Research 21~(118)
  (2020) 1--6.
\newline\urlprefix\url{http://jmlr.org/papers/v21/20-091.html}

\bibitem{Mottonen:2004vly}
M.~M\"ott\"onen, J.~J. Vartiainen, V.~Bergholm, M.~M. Salomaa, {Quantum
  Circuits for General Multiqubit Gates}, Phys. Rev. Lett. 93~(13) (2004)
  130502.
\newblock \href {https://doi.org/10.1103/PhysRevLett.93.130502}
  {\path{doi:10.1103/PhysRevLett.93.130502}}.

\bibitem{Fadol:2022umw}
A.~Fadol, Q.~Sha, Y.~Fang, Z.~Li, S.~Qian, Y.~Xiao, Y.~Zhang, C.~Zhou,
  {Application of quantum machine learning in a Higgs physics study at the
  CEPC}, Int. J. Mod. Phys. A 39~(01) (2024) 2450007.
\newblock \href {http://arxiv.org/abs/2209.12788} {\path{arXiv:2209.12788}},
  \href {https://doi.org/10.1142/S0217751X24500076}
  {\path{doi:10.1142/S0217751X24500076}}.

\bibitem{Bravo-Prieto:2019kld}
C.~Bravo-Prieto, R.~LaRose, M.~Cerezo, Y.~Subasi, L.~Cincio, P.~J. Coles,
  {Variational Quantum Linear Solver}, Quantum 7 (2023) 1188.
\newblock \href {http://arxiv.org/abs/1909.05820} {\path{arXiv:1909.05820}},
  \href {https://doi.org/10.22331/q-2023-11-22-1188}
  {\path{doi:10.22331/q-2023-11-22-1188}}.

\bibitem{Kandala:2017vok}
A.~Kandala, A.~Mezzacapo, K.~Temme, M.~Takita, M.~Brink, J.~M. Chow, J.~M.
  Gambetta, {Hardware-efficient variational quantum eigensolver for small
  molecules and quantum magnets}, Nature 549~(7671) (2017) 242--246.
\newblock \href {http://arxiv.org/abs/1704.05018} {\path{arXiv:1704.05018}},
  \href {https://doi.org/10.1038/nature23879} {\path{doi:10.1038/nature23879}}.

\bibitem{Park:2024rim}
C.-Y. Park, M.~Kang, J.~Huh, {Hardware-efficient ansatz without barren plateaus
  in any depth}, 2024.
\newblock \href {http://arxiv.org/abs/2403.04844} {\path{arXiv:2403.04844}}.

\bibitem{Jones:2019knd}
T.~Jones, A.~Brown, I.~Bush, S.~C. Benjamin, {QuEST and High Performance
  Simulation of Quantum Computers}, Sci. Rep. 9~(1) (2019) 10736.
\newblock \href {https://doi.org/10.1038/s41598-019-47174-9}
  {\path{doi:10.1038/s41598-019-47174-9}}.

\bibitem{Black:2022cth}
K.~M. Black, et~al., {Muon Collider Forum report}, JINST 19~(02) (2024) T02015.
\newblock \href {http://arxiv.org/abs/2209.01318} {\path{arXiv:2209.01318}},
  \href {https://doi.org/10.1088/1748-0221/19/02/T02015}
  {\path{doi:10.1088/1748-0221/19/02/T02015}}.

\bibitem{Accettura:2023ked}
C.~Accettura, et~al., {Towards a muon collider}, Eur. Phys. J. C 83~(9) (2023)
  864, [Erratum: Eur.Phys.J.C 84, 36 (2024)].
\newblock \href {http://arxiv.org/abs/2303.08533} {\path{arXiv:2303.08533}},
  \href {https://doi.org/10.1140/epjc/s10052-023-11889-x}
  {\path{doi:10.1140/epjc/s10052-023-11889-x}}.

\bibitem{Pedregosa:2011ork}
F.~Pedregosa, et~al., {Scikit-learn: Machine Learning in Python}, J. Machine
  Learning Res. 12 (2011) 2825--2830.
\newblock \href {http://arxiv.org/abs/1201.0490} {\path{arXiv:1201.0490}}.

\bibitem{Kane:1984bb}
G.~L. Kane, W.~W. Repko, W.~B. Rolnick, {The Effective W+-, Z0 Approximation
  for High-Energy Collisions}, Phys. Lett. B 148 (1984) 367--372.
\newblock \href {https://doi.org/10.1016/0370-2693(84)90105-9}
  {\path{doi:10.1016/0370-2693(84)90105-9}}.

\bibitem{Boos:1997gw}
E.~Boos, H.~J. He, W.~Kilian, A.~Pukhov, C.~P. Yuan, P.~M. Zerwas, {Strongly
  interacting vector bosons at TeV e+ e- linear colliders}, Phys. Rev. D 57
  (1998) 1553.
\newblock \href {http://arxiv.org/abs/hep-ph/9708310}
  {\path{arXiv:hep-ph/9708310}}, \href
  {https://doi.org/10.1103/PhysRevD.57.1553}
  {\path{doi:10.1103/PhysRevD.57.1553}}.

\bibitem{Ruiz:2021tdt}
R.~Ruiz, A.~Costantini, F.~Maltoni, O.~Mattelaer, {The Effective Vector Boson
  Approximation in high-energy muon collisions}, JHEP 06 (2022) 114.
\newblock \href {http://arxiv.org/abs/2111.02442} {\path{arXiv:2111.02442}},
  \href {https://doi.org/10.1007/JHEP06(2022)114}
  {\path{doi:10.1007/JHEP06(2022)114}}.

\end{thebibliography}
\bibliographystyle{elsarticle-num}

\end{document}